\documentclass[a4paper]{article}
\usepackage{enumerate}
\usepackage{graphicx}
\usepackage{epsfig}

\newcommand{\equ}{\begin{equation}}
\newcommand{\nequ}{\end{equation}}
\newcommand{\no}{\nonumber\\}

\def\f{\frac}
 \newcommand{\Ref}[1]{(\ref{#1})}
 
\begin{document}

\title{\bf{Graviton propagator in loop quantum gravity}} 

\author{\large Eugenio Bianchi${\,}^{a}$,
 Leonardo
 Modesto${\,}^{bc}$, Carlo Rovelli${\,}^c$ and Simone Speziale${\,}^d$
 \\[1mm]
\em\small{${}^a$ Scuola Normale Superiore and INFN, Piazza dei Cavalieri 7, I-56126 Pisa, EU}\\[-1mm]
\em\small{${}^b$ Dipartimento di Fisica, Universit\`a di Bologna,  g Via Irnerio 46, I-40126 Bologna, EU}\\[-1mm]
\em \small{${}^c$ Centre de Physique Th\'eorique de Luminy%
\footnote{Unit\'e mixte de recherche (UMR 6207) du CNRS et des Universit\'es
de Provence (Aix-Marseille I), de la Meditarran\'ee (Aix-Marseille II) et du Sud (Toulon-Var); laboratoire affili\'e \`a la FRUMAM (FR 2291).} , Universit\'e de la M\'editerran\'ee, F-13288 Marseille, EU}
\\[-1mm]
\em \small{${}^d$ Perimeter Institute, 31 Caroline St.\,N, Waterloo, N2L 2Y5, Ontario, Canada.}}

\date{\small\today} 
\maketitle

\begin{abstract}
\noindent  We compute some components of the graviton propagator in loop quantum gravity, using the spinfoam formalism, up to some second order  terms in the expansion parameter.   
\end{abstract}

\section{Introduction}
An open problem in quantum gravity is to compute particle scattering amplitudes from the  full background--independent theory, and recover low--energy physics \cite{qg}. The difficulty is that general covariance makes conventional $n$-point functions ill--defined in the absence of a background.  A strategy for addressing this problem has been suggested in \cite{scattering}; the idea is to study the boundary amplitude, namely the functional integral over a finite spacetime region, seen as a function of the boundary value of the field \cite{oeckl}.  In conventional quantum field theory, this boundary amplitude is well--defined (see \cite{cr,Doplicher:2004gc} ) and codes the physical information of the theory; so does in quantum gravity, but in a fully background--independent manner \cite{cdort}. A generally covariant definition of $n$-point functions can then be based on the idea that the distance between physical points --arguments of the $n$-point function-- is determined by the state of the gravitational field on the boundary of the spacetime region considered.  This strategy was first implemented in the letter \cite{carlo}, where some components of the graviton propagator were computed to the first order in the expansion parameter $\lambda$.  For an implementation of these ideas in 3d, see \cite{simone,SE}.

Here we develop in more detail the calculation presented in  of  \cite{carlo}, and we extend it to terms of second order in  $\lambda$.  We compute a term in the (connected) two-point function, starting from full non-perturbative quantum general relativity, in an appropriate large distance limit.   Only a few components of the boundary states contribute to low order on $\lambda$.  This reduces the model to a 4d generalization of the ``nutshell" 3d model studied in \cite{nutshell}.    The associated boundary amplitude can be read as the creation, interaction and annihilation of few ``atoms of space", in the sense in which Feynman diagrams in conventional quantum field theory expansion can be viewed as creation, interaction and annihilation of particles.   Using a natural gaussian form of the vacuum state, peaked on the intrinsic \emph{as well as the} extrinsic geometry of the boundary, we derive an expression for a component of the graviton propagator. At large distance, this agrees with the conventional graviton propagator.  

Our main motivation is to show that a technique for computing particle scattering amplitudes in background--independent theories can be developed. (The viability of the notion of particle in a finite region is discussed in \cite{daniele}.  For the general relativistic formulation of quantum mechanics underlying this calculation, see \cite{book}.  On the relation between graviton propagator and 3-geometries transition amplitudes in the conventional perturbative expansion, see \cite{Mattei:2005cm}.)    

We consider here riemaniann general relativity without matter.  We use basic loop quantum gravity (LQG) results \cite{lqg,lqg2,eigen2}, and define the dynamics by means of a spinfoam technique (for an introduction see \cite{book,ashtekar,thomas} and \cite{Baez,Perez}).  The specific model we use as example is the theory $GFT/B$, in the terminology of \cite{book}, defined using group field theory methods \cite{DFKR,freidel}. On the definition of spin network states in group field theory formulation of spin foam models, see \cite{Mikovic:2001yg} and \cite{book}.   The result extends immediately also to the theory $GFT/C$. These are background independent spinfoam theories. The first was introduced in \cite{DFKR} and is favored by a number of arguments recently put forward \cite{reasons,Conrady:2005qu}. The second  was introduced in \cite{PR} (see also \cite{PR2}) and is characterized by particularly good finiteness properties \cite{finiteness}.   

The physical correctness of these theories has been questioned because in the large distance limit  their interaction vertex (10$j$ symbol, or Barrett-Crane vertex amplitude \cite{BC}) has been shown to include --beside the ``good" term approximating the exponential of the Einstein-Hilbert action  \cite{BarrettWilliams}-- also two ``bad"  terms: an exponential with opposite sign, giving the cosine of Regge action \cite{BarrettWilliams} (analogous to the cosine in the Ponzano--Regge model) and a dominant term that depends on the existence of degenerate four-simplices \cite{BCE,FL,BS}. We show here that only the ``good" term contributes to the propagator. The others are suppressed by the rapidly oscillating phase in the vacuum state that peaks the state on its correct extrinsic geometry. Thus, the physical state selects the ``forward" propagating \cite{etera} component of the transition amplitude. This phenomenon was anticipated in \cite{colosi}.

\section{The strategy: two-point function from the boundary amplitude}

We begin by illustrating the quantities and some techniques that we are going to use in quantum gravity within a simple context. 

\subsection{A single degree of freedom}

Consider the two-point function of a single harmonic oscillator with mass $m$ and angular frequency $\omega$.  This is given by 
\begin{equation}
G_0(t_1,t_2) = \langle 0 |x(t_1) x(t_2)| 0 \rangle  = \langle 0 |x\ e^{-\frac i\hbar H(t_1-t_2)} \ x| 0 \rangle
\label{tpho}
\end{equation}
where $| 0 \rangle$ is the vacuum state, $x(t)$ is the Heisenberg position operator at time $t$ and $H$ the hamiltonian. We write a subscript $_0$ in $G_0(t_1,t_2)$ to remind us that this is an expectation value computed on the vacuum state. Later we will also consider similar expectation values  computed on other classes of states, as for instance in 
\begin{equation}
G_\psi(t_1,t_2) = \langle \psi |x(t_1) x(t_2)| \psi \rangle.
\label{tphpsi}
\end{equation}
Elementary creation and annihilation operator techniques give 
\begin{equation}
G_0(t_1,t_2) =\frac{\hbar}{2m\omega}\  e^{-\frac32 i \omega (t_1-t_2)}.
\label{result}
\end{equation}
In the Schr\"odinger picture, the r.h.s.\ of   (\ref{tpho}) reads
\begin{equation}
G_0(t_1,t_2) =  \int dx_1dx_2\  \overline{\psi_0(x_1)}\  x_1\ W[x_1,x_2;t_1,t_2]\ x_2\ {\psi_0(x_2)}
 \label{tpho2}
\end{equation}
where $\psi_0(x)=  \langle x |0\rangle$ is the vacuum state and $W[x_1,x_2;t_1,t_2]$ is the propagator, namely the matrix element of the evolution operator 
\begin{equation}
W[x_1,x_2;t_1,t_2]= \langle x_1| e^{-iH(t_1-t_2)}|  x_2 \rangle.
\label{propagatore}
\end{equation}
Recalling that 
\begin{equation}
\psi_0(x) = {}^4\!\!\!\!\sqrt{\frac{m\omega}{\pi\hbar}}\,  e^{-\frac{m\omega}{2\hbar}x^2}
\end{equation}
and (see for instance \cite{book}, page 168 and {\em errata})
\begin{equation}
 W(x_1,x_2;T)=\sqrt{\frac{m\omega}{2\pi i\hbar\sin{\omega T}}}\  e^{i\frac{m\omega}{2\hbar}
 \, \frac{\left(x_1^2+x_2^2\right)\cos\omega T-2x_1x_2}{\sin\omega T}}
\end{equation}
are two gaussian expressions, we obtain the two-point function (\ref{tpho}) as the second momentum of a gaussian 
\begin{equation}
G_0(t_1,t_2) = \frac{m\omega}{\pi\hbar}
\sqrt{\frac{1}{2i\sin{\omega T}}}\,   \int dx_1dx_2\ x_1x_2\  e^{i\frac{m\omega}{2\hbar}
 \, \frac{\left(x_1^2+x_2^2\right)\cos\omega T-2x_1x_2}{\sin\omega T}}\  e^{-\frac{m\omega}{\hbar}x^2}, 
\label{product}
\end{equation}
where the gaussian is the product of a ``bulk" gaussian term and a ``boundary" gaussian term.   Using 
\begin{equation}
\int dx_1dx_2\ x_1x_2\  e^{-\frac{1}{2}(xAx)}=\frac{2\pi}{\sqrt{\det A}}A^{-1}_{12},
\end{equation}
the evaluation of the integral in (\ref{product}) is straightforward. It gives
\begin{equation}
G_0(t_1,t_2) = \frac{m\omega}{\pi\hbar}
\sqrt{\frac{1}{2i\sin{\omega(t_1-t_2)}}}\,  \frac{2\pi}{\sqrt{\det A}}A^{-1}_{12}
\label{W12}
\end{equation}
 in terms of the inverse of the covariance matrix of the gaussian
\begin{equation}
A =\frac{m\omega}{\hbar}\left(\begin{array}{cc} 1- i\frac{\cos\omega(t_1-t_2)}{\sin\omega(t_1-t_2)}   &  \frac{i}{\sin\omega(t_1-t_2)} \\    \frac{i}{\sin\omega(t_1-t_2)}  & 1-i \frac{\cos(t_1-t_2)}{\sin\omega(t_1-t_2)}\end{array}\right)=\frac{-im\omega}{\hbar\sin\omega(t_1-t_2)}\left(\begin{array}{cc} e^{i\omega(t_1-t_2)}  & -1 \\ -1 &  e^{i\omega(t_1-t_2) }\end{array}\right).\label{matrix2}
\end{equation}
The matrix $A$ is easy to invert and (\ref{W12}) gives precisely (\ref{result}). We will find precisely this structure of a similar matrix to invert at the end of the calculation of this paper. 

Notice that the two-point function 
(\ref{tpho}) can also be written as the (analytic continuation of the euclidean version of) the functional integral
\begin{equation}
G_0(t_1,t_2) =  \int Dx(t)\ x(t_1) x(t_2)\ e^{i\int_{-\infty}^{\infty}L(x,dx/dt)}. 
\label{fiho}
\end{equation}
where $L$ is the harmonic oscillator lagrangian, and the measure is appropriately normalized. Let us break the (infinite number of) integration variables $x(t)$ in various groups: those where $t$ is less, equal or larger than, respectively, $t_1$ and $t_2$. Using this, and writing the integration variable $x(t_1)$ as $x_1$ and  the integration variable $x(t_2)$ as $x_2$, we can rewrite (\ref{fiho}) as
\begin{equation}
G_0(t_1,t_2) =  \int dx_1dx_2\ \   \overline{\psi_0(x_1)}\ x_1  \  W[x_1,x_2;t_1,t_2]\ x_2 \  \psi_0(x_2) 
\label{eccoloho}
\label{fiho2}
\end{equation}
where 
\begin{equation}
W[x_1,x_2;t_1,t_2] =  \int_{x(t_2)=x_2}^{x(t_1)=x_1} Dx(t) \ e^{i\int_{t_2}^{t_1}L(x,dx/dt)}
\end{equation}
is the functional integral restricted to the open interval $(t_1,t_2)$ integrated over the paths that start at $x_2$ and end at $x_1$;  while
\begin{equation}
\psi_0(x) =  \int_{x(-\infty)=0}^{x(t_1)=x} Dx(t) \ e^{i\int_{-\infty}^{t_1}L(x,dx/dt)}
\end{equation}
is the functional integral restricted to the interval $(-\infty,t_1)$. As well known, in the euclidean theory this gives the vacuum state. Thus, we recover again the form (\ref{tpho2}) 
of the two-point function, with the additional  information that the ``bulk" propagator term can be viewed as the result of the functional integral in the interior of the $(t_1,t_2)$ interval, while the ``boundary" term can be viewed as the result of the functional integral in the exterior.  In this language the specification of the particular state $| 0 \rangle$ on which the expectation value of $x(t_1) x(t_2)$ is computed, is coded in the boundary behavior of the functional integration variable at infinity: $x(t)\to 0$ for $t\to\pm\infty$.

The normalization of the functional measure in (\ref{fiho}) is determined by
\begin{equation}
1 = \int Dx(t)\ e^{i\int_{-\infty}^{\infty}L(x,dx/dt)}. 
\label{fihoN}
\end{equation}
Breaking this functional integral in the same manner as the above one gives 
\begin{equation}
1 =  \int dx_1dx_2\ \   \overline{\psi_0(x_1)} \  W[x_1,x_2;t_1,t_2] \  \psi_0(x_2) 
\label{fiho2N}
\end{equation}
or equivalently 
\begin{equation}
1   = \langle 0 | e^{-\frac i\hbar H(t_1-t_2)} | 0 \rangle. 
\label{tphoN}
\end{equation}

Let us comment on the interpretation of (\ref{eccoloho}) and (\ref{fiho2N}), since analogues of these equation will play a major role below. Observe that 
(\ref{eccoloho}) can be written in the form
\begin{equation}
G_0(t_1,t_2) =  \langle W_{t_1,t_2}\  |\; \hat x_1\,  \hat x_2 \,  \Psi_{0} \rangle,
\label{boundaryform}
\end{equation}
in terms of states and operators living in the Hilbert space ${\cal K}_{t_1, t_2} ={\cal H}^*_{t_1}\times {\cal H}_{t_2}$ (the tensor product of the space of states at time $t_1$ and the space of states at time $t_2$) formed by functions $\psi(x_1,x_2)$. (See Section 5.1.4 of \cite{book} for details on $ {\cal K}_{t_1,t_2}$.)    Using the relativistic formulation of quantum mechanics developed in \cite{book}, this expression can be directly re-interpreted as follows.
(i) The ``boundary state" $ \Psi_{0}(x_1,x_2)=\overline{\psi_0(x_1)}\psi_0(x_2)$
represents the joint boundary configuration of the system at the two times $t_1$ and $t_2$, if no excitation of the oscillator is present; it describes the joint outcome of a measurement at $t_1$ and a measurement at $t_2$, both of them detecting no excitations. (ii) The two operators  $x_1$ and $x_2$ create a (``incoming") excitation at $t=t_2$ and a (``outgoing") excitation at $t=t_1$; thus the state  $\hat x_1 \hat x_2   \Psi_{0}$ can be interpreted as a boundary state representing the joint outcome of a measurement at $t_1$ and a measurement at $t_2$, both of them detecting a single excitation. 
(iii) The bra $W_{t_1,t_2}(x_1,x_2)=W[x_1,x_2;t_1,t_2]$ is the linear functional coding the dynamics, whose action on the two-excitation state associates it an amplitude, which can be compared with other similar amplitudes.   For instance, observe that 
\begin{equation}
\langle W_{t_1,t_2}\  |\ \hat x_2\   \Psi_{t_1,t_2} \rangle = 0;
\end{equation}
that is, the probability amplitude of measuring a single excitation at $t_2$ and no excitation at $t_1$ is zero. Finally, the normalization condition  (\ref{fiho2N}) reads
\begin{equation}
1= \langle W_{t_1,t_2}\  | \Psi_{0} \rangle;
\label{WdW0}
\end{equation}
which requires that the boundary state $\Psi_{0}$ is a solution of the dynamics, in the sense that its projection on $t_1$ is precisely the time evolution of its projection to  $t_2$. As we shall see below, this condition generalizes to the case of interest for general relativity.  We call
(\ref{WdW0}) the ``Wheeler-deWitt" (WdW) condition. This condition satisfied by the boundary state should not be confused with the normalization condition, 
\begin{equation}
1= \langle  \Psi_{0}  | \Psi_{0} \rangle,
\label{norm0}
\end{equation}
which is also true, and which follows immediately from the fact that $|0\rangle$ is normalized in ${\cal H}_t$.  

In general, given a state $\Psi\in {\cal K}_{t_1,t_2}$,  the equations
\begin{equation}
\langle W_{t_1,t_2}\  | \Psi \rangle= 1;
\label{WdW}
\end{equation}
and
\begin{equation}
\langle  \Psi  | \Psi \rangle = 1,
\label{norm}
\end{equation}
 \emph{are} equivalent to the full quantum dynamics, in the following sense. If the state is of the form $\Psi=\bar\psi_{\rm f}\otimes\psi_{\rm i}$, then (\ref{WdW}) and (\ref{norm})
imply that 
\begin{equation}
\psi_{\rm f}=e^{-iHt}\ \psi_{\rm i}.
\label{dinamica}
\end{equation}
 
Finally, recall that a coherent (semiclassical) state $\psi_{ {\mathbf q}}(x)\sim e^{-\frac{\alpha}2(x-q)^2+\frac i\hbar px}$ is peaked on values $q$ and $p$ of position and momentum.  In particular, the vacuum state of the harmonic oscillator is the coherent state peaked on the values $q=0$ and  $p=0$, with $\alpha=m\omega/\hbar$.  Thus we can write $\psi_0 = \psi_{(q=0,p=0)}$.  In the same manner, the boundary state  $\Psi_{0}=\overline{\psi_0(x_1)}\psi_0(x_2)$ can be viewed as a coherent \emph{boundary} state,  associated with the values $q_1=0$ and  $p_1=0$ at $t_1$ and $q_2=0$ and  $p_2=0$ at $t_2$.  We can write a  \emph{generic} coherent boundary state as 
\begin{equation}
\Psi_{q_1,p_1,q_2,p_2}(x_1,x_2)  =  \overline{\psi_{(q_1,p_1)}(x_1)}\; \psi_{(q_2,p_2)}(x_2).
\label{boundarycoherent1}
\end{equation}
A special case of these coherent boundary states is obtained when $(q_1, p_1)$ are the classical evolution at time $t_1-t_2$ of the initial conditions  $(q_2, p_2)$. That is, when in the $t_1-t_2$ interval there exists a solution  $q(t),p(t)$ of the classical equations of motion precisely bounded by $q_1,p_1,q_2,p_2$, namely such that $q_1=q(t_1), p_1=p(t_1)$ and $q_2=q(t_2),p_2=p(t_2)$. 
If such a classical solution exists, we say that the quadruplet $(q_1,p_1,q_2,p_2)$ is \emph{physical}.   As well known the harmonic oscillator dynamics gives in this case  $e^{-iH(t_1-t_2)}\Psi_{q_2,p_2}=\Psi_{q_1,p_1}$, or 
\begin{equation}
 \langle W_{t_1,t_2}\  | \Psi_{q_1,p_1,q_2,p_2} \rangle = 1. 
\end{equation}
That is, it satisfies the WdW condition (\ref{WdW}).
In this case, we denote the semiclassical boundary state a \emph{physical} semiclassical boundary states. The vacuum boundary state $\Psi_0$ is a particular case of this: it is the physical semiclassical boundary state determined by the classical solution $q(t)=0$ of the equations of motion, which is the one with minimal energy.  
Given a physical boundary state, we can consider a two-point function describing the propagation of a quantum excitation ``over" the semiclassical trajectory $q(t),p(t)$ as
\begin{equation}
G_{q_1,p_1,q_2,p_2}(t_1,t_2) = \langle \psi_{(q_1,p_1)} |x(t_1) x(t_2)| \psi_{(q_2,p_2)} \rangle  =  \langle W_{t_1,t_2}\  |\ \hat x_1 \hat x_2\, \Psi_{q_1,p_1,q_2,p_2} \rangle. 
\label{tphoqpqp}
\end{equation}
This quantity will pay a considerable role below. Indeed, the main idea here is to compute quantum--gravity $n$-point functions using states that describe the boundary value of the gravitatonal field on given boundary surfaces. 

There is an interesting phenomenon regarding the \emph{phases} of the boundary state $\Psi_{q_1,p_1,q_2,p_2}(x_1,x_2)$ and of the propagator 
$W_{t_1,t_2}(x_1,x_2)$ that should be noticed.  If $p_1$ and $p_2$ are different from zero, they give rise to a phase factor $e^{-{i\over\hbar}(p_1x_1-p_2x_2)}$, in the boundary state. In turn, it is easy to see that $W_{t_1,t_2}(x_1,x_2)$ contains precisely the inverse of this same phase factor, when expanded around $(q_1,q_2)$. In fact, the phase of the propagator is the classical Hamilton function $S_{t_1,t_2}(x_1,x_2)$ (the value of the action, as a function of the boundary values \cite{book}). Expanding the Hamilton function around 
$q_1$ and $q_2$ gives to first order
\begin{equation}
S_{t_1,t_2}(x_1,x_2)=S_{t_1,t_2}(q_1,q_2)+{\partial S\over \partial x_1}(x_1-q_1)+{\partial S\over \partial x_2}(x_2-q_2),
\end{equation}
 but 
\begin{equation}
{\partial S\over \partial x_1}=p_1\hspace{1em}{\rm and }\hspace{1em}
{\partial S\over \partial x_2}=-p_2.
\end{equation}
Giving a phase factor $e^{{i\over\hbar}(p_1x_1-ip_2x_2)}$, which is precisely the inverse of the one in the boundary state.
In the Schr\"odinger representation of (\ref{tphoqpqp}), the gaussian factor in the
boundary state peaks the integration around $(q_1,q_2)$; in this region, 
we have that {\em the phase of the boundary state is determined by the classical value of the momentum, and is cancelled by a corresponding phase factor in the propagator $W$.} In particular, the rapidly oscillating phase in the boundary state fails to suppress the integral precisely because it is compensated by a corresponding rapidly oscillating phase in $W$.  This, of course, is nothing that the realization, in this language, of the well--known emergence of classical trajectories from the constructive coherence of the quantum amplitudes.  This phenomenon, noted in \cite{carlo} in the context of quantum gravity, plays a major role below.

\subsection{Field theory}

Let us now go over to field theory. The two-point function (or particle propagator) is defined by the (analytic continuation of the  euclidean version of the) path integral ($\hbar=1$ from now on)
\begin{equation}
G_0(x,y) =\langle 0 |\phi(x) \phi(y)| 0 \rangle  =  \langle 0 |\phi(\vec x)\ e^{-iH(x_0-y_0)} \ \phi(\vec y)| 0 \rangle=  \int D\phi(x)\ \phi(x) \phi(y)\ e^{iS[\phi]},
\label{fifi}
\end{equation}
where the normalization of the measure is determined by 
\begin{equation}
1 =  \int D\phi(x)\ \ e^{iS[\phi]}
\label{fifiN}
\end{equation}
and the $_0$ subscript reminds that these are expectation values of products of field operators in the particular state $| 0 \rangle $.
These equations generalize (\ref{fiho}) and (\ref{fihoN})  to field theory.%
\footnote{A well-known source of confusion is of course given by the fact that in the case of a free particle the propagator (\ref{propagatore}) coincides with the 2-point function of the free field theory.}
   As before, we can break the integration variables of the path integral in various groups.  For instance,
in the values of the field in the five spacetime--regions identified by $t$ being less, equal or larger than, respectively, $ x_0$ and $y_0$. This gives a Schr\"odinger representation of the two-point function of the form 
\begin{equation}
G_0(x,y) =   \int D\varphi_1 D\varphi_2\  \overline{\psi_0(\varphi_1)}\  \varphi_1(\vec x)\ W[\varphi_1,\varphi_2;(x_0-y_0)]\ \varphi_2(\vec y)\ {\psi_0(\varphi_2)}. 
\label{fifree}
\end{equation}
where $\varphi_1$ is the three-dimensional field at time $t_1$, and
$\varphi_2$ is the three-dimensional field at time $t_2$.  For a free field, 
the field propagator (or propagation kernel)
\begin{equation}
W(\varphi_1,\varphi_2;T)= \langle \varphi_1| e^{-iHT}|  \varphi_2 \rangle.
\end{equation}
and the boundary vacuum state are gaussian expression in the boundary field $\varphi=(\varphi_1,\varphi_2)$. These expressions, and the functional integral (\ref{fifree}), are explicitly computed in \cite{Doplicher:2004gc}. 
In a free theory, the boundary vacuum state can be written as a physical semiclassical state peaked on vanishing field and momentum $\pi$, as in  (\ref{boundarycoherent1}):
\begin{equation}
\Psi_{0}(\varphi_1,\varphi_2) \equiv 
\Psi_{\varphi_1=0,\pi_1=0,\varphi_2=0,\pi_2=0}(\varphi_1,\varphi_2)
 =  
\overline{\psi_{0}(\varphi_1)}\; \psi_{0}(\varphi_2) .  
\label{boundarycoherentfi}
\end{equation}
Notice that the momentum $\pi=\frac{d\varphi_1}{dt}$ is the derivative of the classical field normal to $\Sigma$.

More interesting for what follows, we can choose a compact finite region $\cal R$ in spacetime, bounded by a closed 3d surface $\Sigma$, such that the two points $x$ and $y$ lie on $\Sigma$. Then we can separate the integration variables in (\ref{fifi}) into those inside $\cal R$, those on 
 $\Sigma$ and  those outside $\cal R$, and thus write the two-point function (\ref{fifi}) in the form \begin{equation}
G_0(x,y) =   \int D\varphi \   \varphi(x)\ \varphi(y)\  W[\varphi;\Sigma]\ {\Psi_{0}(\varphi)},
\label{eccolo2}
\end{equation}
where $\varphi$ is the field on $\Sigma$,
\begin{equation}
W[\varphi;\Sigma]\  =  \int_{\partial\phi=\varphi} D\phi_R  \  e^{-iS_{\cal R}[\phi_R]}
\label{eccolo3}
\end{equation}
is the functional integral restricted to the region $\cal R$, and integrated over the interior fields $\phi_R$ bounded by the given boundary field $\varphi$. The boundary state  ${\Psi_0(\varphi)}$ is given by the integral restricted to the outside region, $\overline{\cal R}$.  The boundary conditions on the functional integration variable
\begin{equation}
\phi_{\overline{\cal R}}(x)\to 0, \hspace{2em}{\rm for }\hspace{2em} |x| \to\infty
\label{bc}
\end{equation}
determine the vacuum state.  In a free theory, this is still a gaussian expression in $\varphi$, but the covariance matrix is non--trivial and is determined by the shape of $\Sigma$. The state $\Psi_0$ can nevertheless be still viewed as a semiclassical boundary state 
associated to the compact boundary, peaked on the value $\varphi=0$ of the field and the value $\pi=0$ of a (generalized) momentum (the derivative of the field normal to the surface) \cite{book}.    Equation (\ref{eccolo2}) will be our main tool in the following. 

In analogy with (\ref{boundaryform}), equation (\ref{eccolo2})  can be written in the form
\begin{equation}
G_0(x,y) =  \langle W_\Sigma \;  |\;  \hat \varphi(x)\, \hat\varphi(y) \;  \Psi_0 \rangle. 
\end{equation}
in terms of states and operators living in a boundary Hilbert space ${\cal K}_\Sigma$ associated with the 3d surface $\Sigma$.  In terms of the relativistic formulation of quantum mechanics developed in \cite{book}, this expression can be interpreted as follows. (i) The ``boundary state" $ \Psi_0$ represents the boundary configuration of a quantum field on a surface $\Sigma$, when no particles are present; it represents the joint outcome of measurements on the entire surface $\Sigma$, showing no presence of particles.
(ii)  The two operators  $\hat \varphi(x)\ \hat\varphi(y) $ create a (``incoming") particle at $y$ and a (``outgoing") particle at $x$; so that the boundary state $ \varphi(x)\, \varphi(y) \,  \Psi_0$ represents the  joint outcome of measurements on $\Sigma$, detecting a (``incoming") particle at 
$y$ and a (``outgoing") particle at $x$. (iii) Finally, the bra $W_\Sigma$ is the linear functional coding the dynamics, whose action on the two-particle boundary state associates it an amplitude, which can be compared with other analogous amplitudes. The normalization condition for the measure, equation (\ref{fifiN}), becomes the WdW condition
 \begin{equation}
1 =  \langle W_\Sigma \  |  \Psi_0 \rangle,
\end{equation}
which singles out the physical boundary states. 

Finally, as before, let $\mathbf q = (q,p)$ be a given couple of boundary values of the field $\varphi$ and its generalized momentum on $\Sigma$.  If there exists a classical solution $\phi$ of the  equations of motion whose restriction to $\Sigma$ is $q$ and whose normal derivative to $\Sigma$ is $p$, then we say that  $\mathbf q =(q,p)$ are \emph{physical} boundary data. Let  $\Psi_{\mathbf q}$ be a boundary state in ${\cal K}_\Sigma$ peaked on these values: schematically 
\begin{equation}
\Psi_{\mathbf q}(\varphi)\sim e^{-\int(\varphi-q)^2+i\int p\phi} .
\label{phasephi}
\end{equation}
  If $\mathbf q=(q,p)$ are physical boundary data, we say that $\Psi_{\mathbf q}$ is a \emph{physical} semiclassical state.  In this case, we can consider the two-point function
\begin{equation}
G_{\mathbf q}(x,y) =  \langle W_\Sigma \;  |\;  \hat \varphi(x)\, \hat\varphi(y) \;  \Psi_{\mathbf q} \rangle
\end{equation}
describing the propagation of a quantum, from $y$ to $x$, over the classical field configuration $\phi$ giving the boundary data $\mathbf q=(q,p)$.  In the Schr\"odinger representation of this
expression, there is a cancellation of the phase of the boundary state  $\Psi_{\mathbf q}$ with the phase of the propagation kernel $W_\Sigma$, analogous to the one we have seen in the case of a single degree of freedom.

\subsection{Quantum gravity}

Let us formally write (\ref{eccolo2}) for pure general relativity, ignoring for the moment problems such as the definition of the integration measure, or ultraviolet divergences.  Given a surface $\Sigma$, we can choose a generalized temporal gauge in which the degrees of freedom of gravity are expressed by the 3-metric $\gamma$ induced on $\Sigma$, with components $\gamma_{ab}(x)\ a,b=1,2,3$. That is, if the surface is locally given by $x^4=0$, we gauge fix the 4d gravitational metric field $g_{\mu\nu}(x)$ by $g_{44}=1, g_{40}=0$, and $\gamma_{ab}=g_{ab}$.  Then the graviton two-point function  (\ref{eccolo2}) reads in this gauge
\begin{equation}
G_0^{abcd}(x,y) =  \int [D\gamma] \  h^{ab}(x)\ h^{cd}(y)\  W[\gamma;\Sigma]\ {\Psi_{0}(\gamma)},
\label{eccolo22}
\end{equation}
where  $h^{ab}(x)= \gamma^{ab}(x)-\delta^{ab}$.   As observed for instance in \cite{cdort}, if we assume that $W[\gamma;\Sigma]$ is given by a functional integration on the bulk, 
as in (\ref{eccolo3}), where measure and action are generally covariant, then we have immediately that $W[\gamma;\Sigma]$ is independent from (smooth deformations of) $\Sigma$. Hence, at fixed topology of $\Sigma$ (say, the surface of a 3-sphere), we have $W[\gamma;\Sigma]=W[\gamma]$, that is 
\begin{equation}
G_0^{abcd}(x,y) =  \int [D\gamma] \  h^{ab}(x)\ h^{cd}(y)\  W[\gamma]\ {\Psi_{0}(\gamma)}. 
\label{eccolo222}
\end{equation}
What is the interpretation of the boundary state ${\Psi_{0}(\gamma)}$ in a general covariant theory? In the case of the harmonic oscillator, the vacuum state $ | 0 \rangle$ is the state that  minimizes the energy. In the case of a free theory on a background, in addition, 
it is the sole Poincar\'e invariant state. In both cases the vacuum state can also be obtained from a functional integral by fixing the behavior of the fields at infinity.
But in background--independent quantum gravity, there is no energy to minimize and no global Poincar\'e invariance.     Furthermore, there is no background metric with respect to which to demand the gravitational field to vanish at infinity. 
In fact,  it is well known that the unicity and the very definition of the vacuum state is highly problematic in nonperturbative quantum gravity (see for instance \cite{book}), a phenomenon that begins to manifest itself already in QFT on a curved background. Thus, in quantum gravity there is a multiplicity of possible states that we can consider as boundary states, instead of a single preferred one.  

Linearized quantum gravity gives us a crucial hint, and provides us with a straightforward way to \emph{interpret} semiclassical boundary states.  Indeed, consider linearized quantum gravity, namely the well--defined theory of a noninteracting spin--2 graviton field $h_{\mu\nu}(x)$ on a flat spacetime with background metric $g^{0}_{\mu\nu}$. This theory has a preferred vacuum state $ | 0\rangle $.    Now, choose a boundary surface $\Sigma$ and denote $\mathbf q=(q,p)$ its three-geometry, formed by the 3-metric $q_{ab}$ and extrinsic curvature field $p^{ab}$, induced on $\Sigma$ by the flat background metric of spacetime.  The vacuum state defines a gaussian boundary state on $\Sigma$,  peaked around $h=0$. We can schematically write this state as $\Psi_\Sigma(h)\sim e^{-\int h^2}$. (In the conventional case in which $\Sigma$ is formed by two parallel hyper-planes, the explicit form of this state is given in \cite{Mattei:2005cm}.) Now, on $\Sigma$ there are two metrics: the metric $q$ induced by the background spacetime metric, and the metric $\gamma=q+h$, induced by the true physical metric $g_{\mu\nu} = g^{0}_{\mu\nu}+h_{\mu\nu}$, which is the sum of the background metric and the dynamical linearized gravitational field.  Therefore the vacuum functional $\Psi_0(h)$ defines  a functional $ \Psi_{ {\mathbf q}}(\gamma)$ {\em of the physical metric $\gamma$ of $\Sigma$} as follows
 \begin{equation}
\Psi_{\mathbf q}(\gamma)= \Psi_{ {\mathbf q}}(q+h)\equiv \Psi_0(h).
 \end{equation}
Schematically 
 \begin{equation}
\Psi_{\mathbf q}(\gamma)= \Psi_0(h)=  \Psi_0(\gamma-q)\sim  e^{-\int(\gamma-q)^2}.
 \end{equation}
A bit more precisely, as was pointed out in \cite{carlo}, we must also take into account a phase term, generated by the fact that the normal derivative of the induced metric does not vanish ($q$ changes if we deform $\Sigma$). This gives, again very schematically 
 \begin{equation}
\Psi_{\mathbf q}(\gamma)\sim  e^{-\int(\gamma-q)^2+i\int p\gamma}
 \end{equation}
as in (\ref{phasephi}). 
Recall indeed that in general relativity the intrinsic and extrinsic geometry play the role of canonical variable and conjugate variable. As pointed out in \cite{carlo}, a semiclassical boundary state must be peaked on both quantities, as coherent states of the harmonic oscillator are equally peaked on $q$ and $p$. The functional $\Psi_{ {\mathbf q}}$ of the metric can immediately be interpreted as a boundary state of quantum gravity, as determined by the linearized theory. Observe that it depends on the background geometry of $\Sigma$, because $q$ and $p$ do: the form of this state is determined by the location of $\Sigma$ with respect to the background metric of space.  Therefore (when seen as a function of the true metric $\gamma$) there are different possible boundary states in the linearized theory, depending on where is the boundary surface. Equivalently, there are different boundary states depending on what is the mean boundary geometry $q$ on $\Sigma$.   

Now, in full quantum gravity we must expect, accordingly, to have many possible distinct semiclassical boundary states $\Psi_{\mathbf q}(\gamma)$ that are peaked on distinct 3-geometries $\mathbf q=(q,p)$. In the background-independent theory they cannot be \emph{anymore} interpreted as determined by the location of $\Sigma$ with respect to the background (because there is no background!).  But they can \emph{still} be interpreted as determined by the mean boundary geometry $ {\mathbf q}$ on $\Sigma$.  Their interpretation is therefore immediate: they represent coherent semiclassical states of the boundary geometry.  The multiplicity of the possible locations of $\Sigma$ with respect to the background geometry in the background-dependent theory, translates into a multiplicity of possible coherent boundary states in the background-independent formalism. 
 
In fact, this conclusion follows immediately from the core physical assumption of general relativity:  the identification of the gravitational field with the spacetime metric.  A coherent boundary state of the gravitational field is peaked, in particular, on a given classical value of the metric. In the background-dependent picture, this can be interpreted as information about the location of $\Sigma$ in spacetime.  In a background-independent picture, there is no location in spacetime: the geometrical properties of anything is solely determined by the local value of the gravitational field.  In a background-independent theory, the dependence on a boundary geometry is not in the location of $\Sigma$ with respect to a background geometry, but rather in the boundary state of the gravitation field on the surface $\Sigma$ itself. 

Having understood this, it is clear that the two-point function of a background-independent theory can be \emph{defined} as a function of the mean boundary geometry, instead of a function of the background metric.  If $\mathbf q=(q,p)$ is a given geometry  of a closed surface $\Sigma$ with the topology of a 3-sphere, and $\Psi_{\mathbf q}$ is a coherent state peaked on this geometry, consider the expression
\begin{equation}
G_{\mathbf q}^{abcd}(x,y) =  \int [D\gamma] \  h^{ab}(x)\ h^{cd}(y)\  W[\gamma]\ {\Psi_{\mathbf q}(\gamma)}. 
\label{eccolo44}
\end{equation}
At first sight, this expression appears to be meaningless. The r.h.s. is completely independent from the location of $\Sigma$ on the spacetime manifold. What is then the meaning of the 4d coordinates $x$ and $y$ in the l.h.s.?   In fact, this is nothing than the usual well--known problem of the conventional definition of $n$-point functions in generally covariant theories: if action and measure are generally covariant, equation (\ref{fifi}) is independent from $x$ and $y$ (as long as $x\ne y$); because a diffeomorphism on the integration variable can change $x$ and $y$, leaving all the rest invariant. We seem to have hit the usual stumbling block that makes $n$-point functions useless in generally covariant theories. 

In fact, we have not, because the very dependence of $G _{ {\mathbf q}}^{abcd}(x,y)$ on $ {\mathbf q}$ provides the obvious solution to this problem: let us \emph{define} a ``generally covariant 2-point function" ${\mathbf G} _{\mathbf q}^{abcd}(x,y)$ as follows. Given a three-manifold $S_3$ with the topology of a 3-sphere, equipped with given fields ${\mathbf q}=(q_{ab}(\mathbf x),p^{ab}(\mathbf y))$, and given two points $\mathbf x$ and $\mathbf y$ \emph{on this metric manifold}, we define 
\begin{equation}
{\mathbf G} _{\mathbf q}^{abcd}(\mathbf x,\mathbf y) =  \int [D\gamma] \  h^{ab}(\mathbf x)\ h^{cd}(\mathbf y)\  W[\gamma]\ {\Psi_{\mathbf q}(\gamma)}. 
\label{eccolo4}
\end{equation}
The difference between (\ref{eccolo44}) and (\ref{eccolo4}) is that in the first expression $x$ and $y$ are coordinates in the background 4d \emph{differential} manifold, while in the second $\mathbf x$ and $\mathbf y$ are points in the 3d \emph{metric} manifold $(S_3, q)$.   It is clear that with this definition the dependence of the 2-point function on $\mathbf x$ and $\mathbf y$ is non trivial: metric relations between $\mathbf x$ and $\mathbf y$ are determined by $ {\mathbf q}$.    In particular, a 3d active diffeomorphism on the integration variable $g$ changes $\mathbf x$ and $\mathbf y$, but also $ {\mathbf q}$, leaving the metric relations between $\mathbf x$ and $\mathbf y$ invariant.  

The physically interesting case is when $\mathbf q=(q,p)$ are a set of \emph{physical} boundary conditions.  Since we are considering here pure general relativity without matter, this means that there exists a Ricci flat spacetime with 4d metric $g$ and an imbedding $\Sigma: S_3\to M$, such that $g$ induces the three metric $q$ and the extrinsic curvature $p$ on $S_3$.  In this case, the semiclassical boundary state $\Psi _{ {\mathbf q}}$ is a physical state. Measure and boundary states must be normalized in such a way that 
\begin{equation}
 \int [D\gamma] \  W[\gamma]\ {\Psi_{\mathbf q}(\gamma)} = 1. 
\label{normalizzazione}
\end{equation}
 Then the two point function  (\ref{eccolo4}) is a non-trivial and invariant function of the physical 4d distance 
\begin{equation}
L = d_{g}(\Sigma(\mathbf x),\Sigma( \mathbf y)).
\label{Lg4}
\end{equation}
It is clear that if $g$ is the flat metric this function must reduce immediately to the conventional 2-point function of the linearized theory, in the appropriate large distance limit.  

This is the definition of a generally covariant two-point function  proposed in \cite{scattering}, which we use here. 

Finally, the physical interpretation of  (\ref{eccolo4}) is transparent: it defines an amplitude associated to a joint set of measurements performed on a surface $\Sigma$ bounding a finite spacetime region, where the measurements include: (i) the average geometry of $\Sigma$ itself, namely the physical distance between detectors, the time lapse between measurements, and so on; as well as (ii) the detection of a (``outgoing") particle (a graviton) at $x$ and the detection of a (``incoming") particle (a graviton) at $y$. The two kinds of measurements, that are considered of different nature in non-generally-relativistic physics, are on equal footing in general relativistic physics (see \cite{book}, pg.\ 152-153).  In generally covariant quantum field theory, the single boundary state $  \hat h^{ab}( x)\hat h^{cd}( y)  \Psi _{ {\mathbf q}} $ codes the two.  Notice that the quantum geometry in the \emph{interior} of the region $\cal R$ is free to fluctuate. In fact, $W$ can be interpreted as the sum over all interior 4-geometries.  What is determined is a boundary geometry as measured by the physical apparatus that surrounds a potential interaction region. 

Equation (\ref{eccolo4}) can be realized concretely in LQG by identifying (i) the boundary Hilbert space associated to $\Sigma$ with the (separable \cite{winston}) Hilbert space spanned by the (abstract) spin  network states $s$, namely the $s$-knot states; (ii) the linearized gravitational field operators $ \hat h^{ab}( x)$ and $\hat h^{cd}( y) $ with the corresponding LQG operators; (iii) the boundary state $\Psi _{ {\mathbf q}}$ with a suitable spin network functional $\Psi _{ {\mathbf q}}[s]$ peaked on the geometry $q$; and finally, (iv) the boundary functional $W[s]$, representing the functional integral on the interior geometries bounded by the boundary geometry $s$, with the $W[s]$ defined by a spin foam model. This, indeed, is given by a sum over interior spinfoams, interpreted as quantized geometries.  This gives the expression
\begin{equation}
{ \mathbf G} _{ {\mathbf q}}^{a b c d}(\mathbf x, \mathbf y) = 
 \sum_s  \ W[s] \  \hat h^{ab}(\mathbf x)\;  \hat h^{cd}( \mathbf y) 
\; \Psi _{ {\mathbf q}}[s].
 \label{simplypropaga1}
\end{equation}
which we analyze in detail in rest of the paper. 
 The WdW condition reads  
\begin{equation}
1 =  \sum_s  \ W[s] \   \Psi _{ {\mathbf q}}[s].
 \label{normalization}
\end{equation}
Using these two equations together, we can write
\begin{equation}
{ \mathbf G} _{ {\mathbf q}}^{a b c d}(\mathbf x, \mathbf y) = 
\frac{ \sum_s  \ W[s] \  \hat h^{ab}(\mathbf x)\;  \hat h^{cd}( \mathbf y) 
\; \Psi _{ {\mathbf q}}[s]}{ \sum_s  \ W[s] \   \Psi _{ {\mathbf q}}[s]},
 \label{prodi}
\end{equation}
a form that allows us to disregard the overall normalization of $W$ and $\Psi _{\mathbf q}$.
We analyze these ingredients in detail in the next section. 

\section{Graviton propagator: definition and ingredients}

Equation (\ref{simplypropaga1}) is well-defined if we choose a dynamical model giving $W[s]$, a boundary state $\Psi _{ {\mathbf q}}[s]$ and a form for the operator $\hat h^{ab}(x)$. In the exploratory spirit of 
\cite{scattering}, we make here some tentative choices for these ingredient.  In particular, we choose the boundary functional $W[s]$ defined by the group field theory $GFT/B$.
We consider here only some lowest order terms in the expansion of $W[s]$ in the $GFT$ coupling constant $\lambda$. Furthermore, we consider only the first order in a large distance expansion.  Our aim is to recover the 2-point function of the linearized theory, namely the graviton propagator,  in this limit.

\subsection{The boundary functional $W[s]$}

We recall the definition of $W[s]$ in the context of the spinfoam theory $GFT/B$,  referring to \cite{book} and \cite{Perez} for motivations and details.  We follow the notation of 
 \cite{book}.  The theory is defined for a field $\phi:(SO(4))^4\to R$ by an action of the form
\begin{equation} 
S[\phi] = S_{\rm kin}[\phi]  + \frac{\lambda}{5!}\  S_{\rm int}[\phi]. 
\end{equation} 
The field $\phi$ can be expanded in modes 
$\phi^{j_1\dots j_4}_{\alpha_1... \alpha_4\, i}$ (see eq.\,(9.71) of  \cite{book}). 
Notation is as follows. The indices $j_n,\ n=1,...,4$ label simple $SO(4)$ irreducible representations. Recall that the irreducible  representations of $SO(4)$ are labelled by 
a pair of spins $(j_+,j_-)$, corresponding to the split of $so(4)=su(2)\times su(2)$ into its self-dual 
and antiself-dual rotations; the simple representations  are the ones for which $j_+=j_-\equiv j$, and are therefore labelled by a single spin $j$.  The index $\alpha_n$ labels the components of vectors in the representation $j_n$. The index $i$ labels an orthonormal basis of intertwiners  (invariant vectors) on the tensor product of the four representations $j_n$.  We choose a basis in which one of the basis elements is the Barrett-Crane intertwiner $i_{BC}$, given in 
eq.\,(9.99) of  \cite{book}.   Expanded in terms of these modes, the kinetic term of the action is  (eq.\,(9.73) of  \cite{book})
\begin{equation} 
    S_{\rm kin}=\frac{1}{2}\sum\limits_{\alpha_n, j_n,i}
   \phi^{j_n}_{\alpha_n i}\ 
   \phi^{j_n}_{\alpha_n i} 
   \label{kin}
\end{equation} 
The interaction term is  (eq.\,(9.74) of  \cite{book})
\begin{equation}\label{vph}
 S_{\rm int}=   \sum_{\alpha_{nm}, j_{nm},i_n}\ 
 {\textstyle\left(\prod_{m}\ 
\phi^{j_{nm}}_{\alpha_{nm},i_n} \right)}
\ {\mathcal{A}}(j_{nm},i_n). 
\end{equation} 
Here the notation is as follows. The indices $n$ and $m$ run from 1 to 5, with  $n\ne m$. $N_{nm}\equiv N_{mn}$ and $\phi^{j_{1m}}_{\alpha_{1m},i_1}=\phi^{j_{12}j_{13}j_{14} j_{15}}_{\alpha_{12}\alpha_{13}\alpha_{14}\alpha_{15},i_1}$ and so on. 
${\mathcal{A}}(j_{nm},i_n)$ is the Barrett-Crane vertex amplitude.  This is 
\begin{equation} \label{B2}
{\mathcal{A}}(j_{nm},i_n) = \left(\prod_n\delta_{i_ni_{\rm BC}}\right)
{\mathcal{B}}(j_{nm}),
\end{equation}
where ${\mathcal{B}}(j_{nm})\equiv{\mathcal{A}}(j_{nm},i_{\rm BC})$ 
is the 10$j$ symbol, given in eq.\,(9.102) of  \cite{book}.  In the following we use also the formal notation $ \int \phi^2 \equiv S_{\rm kin}[\phi]$ and $\int \phi^5\equiv S_{\rm int}[\phi]$. 

 $SO(4)$-invariant observables of the theory are computed as the expectation values
\begin{equation}
W[s]  = \frac1Z\int \mathrm{D} \phi \ \ f_s(\phi) \  \  \mbox{e}^{-\int \phi^2 - \frac{\lambda}{5!} \int \phi^5}  
\label{ev}
\end{equation}
where the normalization $Z$ is the functional integral without $f_s(\phi)$,
and $f_s(\phi)$ is the function of the field determined by the spin network $s=(\Gamma,j_l,i_n)$.
Recall that a spin network is a graph $\Gamma$ formed by nodes $n$ connected by links $l$, colored with representations $j_l$ associated to the links and intertwiners $i_n$ associated to the nodes.    We note $l_{nm}$ a link connecting the nodes $n$ and $m$, and $j_{nm}\equiv j_{mn}$ the corresponding color.  The spin network function is defined in terms
of the modes introduced above by 
\begin{equation}
f_s(\phi)= \sum_{\alpha_{nm}} \prod_n \phi^{j_{nm}}_{\alpha_{nm}i_n}.
\end{equation}
Here $n$ runs over the nodes and, for each $n$, the index $m$ runs over the four nodes that bound the four links $l_{nm}$ joining at $n$. Notice that each index $\alpha_{nm} \equiv \alpha_{mn}$ appears exactly twice in the sum, and are thus contracted.

Fixed a spin network $s$, (\ref{ev}) can be treated by a perturbative expansion
in $\lambda$, which leads to a sum over Feynman diagrams. Expanding both numerator and denominator,
we have
\begin{eqnarray}
W[s]  &=& \frac1{Z_0}  \int \mathrm{D} \phi \, f_s(\phi) \, \mbox{e}^{-\!\int\! \phi^2}  -  \\ \nonumber &&+ \frac1{Z_0} \frac{\lambda}{5!} \left[\int \mathrm{D} \phi \, f_s(\phi) \left(\!\int \phi^5\!\right) \mbox{e}^{-\!\int\! \phi^2} -\frac{\int \mathrm{D} \phi \, \left(\!\int \phi^5\!\right) \mbox{e}^{-\!\int\! \phi^2}}{Z_0} \int \mathrm{D} \phi \, f_s(\phi) \, \mbox{e}^{-\!\int\! \phi^2}\right]+   \\ \nonumber &&+ \f1{Z_0}\frac{\lambda^2}{2 (5!)^2}\left[ \int \mathrm{D} \phi \, f_s(\phi) \, \left(\!\int \phi^5\! \right)^2 \, \mbox{e}^{-\!\int\! \phi^2}+\ldots\right],
\label{diliberto}
\end{eqnarray}
where
$Z_0=\int \mathrm{D} \phi \, \mbox{e}^{-\!\int\! \phi^2}$.
As usual in QFT, the normalization $Z$ gives rise to all vacuum--vacuum transition amplitudes, and it role is to eliminate disconnected diagrams.

Recall that this Feynman sum  can be expressed as a sum over all connected spinfoams $\sigma=(\Sigma,j_f,i_e)$ bounded by the spin network $s$.   A spinfoam is a two-complex $\Sigma$, namely an ensemble of faces $f$ bounded by edges $e$, in turn bounded by vertices $v$, colored with representations $j_f$ associated to the faces and intertwiners $i_e$ associated to the edges. 

The boundary of a spinfoam $\sigma=(\Sigma,j_f,i_e)$ is a spin network  $s=(\Gamma,j_l,i_n)$, where the graph $\Gamma$ is the boundary of the two-complex $\Sigma$, $j_l=f_f$ anytime the link $l$ of the spin network bounds a face $f$ of the spinfoam and $i_n=i_e$ anytime the node $n$ of the spin network bounds an edge $e$ of the spinfoam. See the Table  \ref{Terminology} for a summary of the terminology. 

\begin{table}[t]\caption{Terminology}\begin{center}\small\begin{tabular}{lccccc}    \hline    \hline    & 0d & 1d & 2d & 3d & 4d\\    \hline &&&&& \\    Spin networks:\  & \emph{node,} &\emph{link;}& & & \\ &&&&&\\    Spinfoams: & \emph{vertex,}\ \  & \emph{edge,}& \emph{face;}& &     \\&&&&& \\    Triangulation: & \emph{point,} & \emph{segment,}\ \ & \emph{triangle,}     & \emph{tetrahedron,} &  \emph{four-simplex.}\   \\ &&&&& \\     \hline      \hline  \end{tabular}\end{center}\label{Terminology}\index{node}\index{link}\index{vertex}\index{edge}\index{face}\index{four-simplex}\end{table}
The amplitudes can be reconstructed from the following Feynman rules; the propagator
\begin{equation}
\mathcal{P}_{\alpha_ni}^{j_n}\,{}_{\alpha'_ni'}^{j'_n}= 
\delta_{i, i'}\  \sum_{\pi(n)}\
\prod_n\  \delta_{j_n, j'\!\!{}_{\pi(n)}}\  
\delta_{\alpha{}_{n}\alpha'\!\!{}_{\pi(n)}} 
    \label{pph}
\end{equation}
where $\pi(n)$ are the permutations of the four numbers $n=1,2,3,4$;  
and the vertex amplitude
\begin{equation}
    \label{vertex}
\mathcal{V}^{\alpha_{nm}i_n}_{j_{nm}}= \Big(\prod_{n}
\delta_{i_{n}i_{\rm BC}}\Big)
\Big(\prod_{n\ne m}
\delta_{\alpha_{nm}\alpha_{mn}}\Big)
\ {\mathcal{B}}(j_{nm}),
\end{equation}
where the index $n=1,...,5 $ labels the five legs of the five-valent vertex; while the index $m\ne n$ labels the four indices on each leg. 

A Feynman graph has vertices $v$ and propagators that we call ``edges" and denote $e$.  A spinfoams $\sigma$ is obtained from a Feynman graph by: (i) selecting one term in each sum over representations and one term in each sum over permutations (eq.\,(\ref{pph})), in the sum that gives the amplitude of the graph; (ii) contracting the closed sequences of $\delta_{\alpha_n\alpha_m}$ in the propagators, vertices and boundary spin-network function; and (iii) associating a \emph{face} $f$, colored by the corresponding representation $j_f$, to each such sequence of propagators and boundary links. See \cite{book} for more details.   We obtain in this manner the amplitude
\begin{equation}
W[s]  = \f1Z \sum_{\sigma, \partial\sigma=s}\ \prod_{f\in\sigma} {\rm dim}(j_f)\  \prod_{v\in\sigma} \lambda {\mathcal{B}}(j^v_{nm}) \left( \prod_{n\in s} \langle i_n  |i_{\rm BC}\rangle   \right).
\label{ssf}
\end{equation}
Here $\sigma$ are spinfoams with vertices $v$ dual to a four--simplex, bounded the spin network $s$.   $f$ are the faces of $\sigma$;  the spins $j^v_{nm}$ label the representations associated to the ten faces adjacent to the vertex $v$, where $n\ne m =1,...,5$; ${\rm dim}(j)$ is the dimension of the representation $j$. The colors of a faces $f$ of $\sigma$ bounded by a link $l$ of $s$ is restricted to match the color of the link: $j_f=j_l$.   The expression is written for arbitrary boundary 
spin-network intertwiners $i_n$: the scalar product is in the intertwiner space and derives from the fact that the vertex amplitude projects on the sole Barrett-Crane intertwiner. The relation between the different elements is summarized in Table \ref{Table2}.  

The sum (\ref{ssf}) can be written as a power series in $\lambda$
\begin{equation}
W[s]=  \sum_{k=0}^\infty \lambda^k\ W_k[s]
\end{equation}
with 
\begin{equation}
W_k[s]= \f1Z \sum_{\sigma^k, \partial\sigma^k=s}
\ \prod_{f\in\sigma} {\rm dim}(j_f)\  \prod_{v\in\sigma} {\mathcal{B}}(j^v_{nm}) \left( \prod_{n\in s} \langle i_n  |i_{\rm BC}\rangle   \right),
\end{equation}
where $\sigma^k$ is a spinfoam with $k$ vertices.

  \begin{table}[ht]    \caption{Relation between a triangulation and    its dual, in the  and 4d bulk and in its 3d boundary.  In parenthesis: adjacent elements.  In    italic, the two-complex and the spin-network's graph. The spinfoam is $\sigma=(\Delta^{*}_{4},j_f,i_e)$. The spin network is $s=(\Delta^{*}_{3},j_l,i_n )$.  }    \begin{center}\small      \begin{tabular}{llll}        \hline        \hline	$\Delta_{4}$ & $\Delta^{*}_{4}$ &&coloring \\        \hline         4-simplex & {\em vertex} &{\footnotesize (5 edg, 	 10 fac)} & \\        tetrahedron & {\em edge}& {\footnotesize 	(4 faces)} & $i_e$   \\        triangle & \em face && $j_f$  \\        segment &  & \\	point & & & \\	\hline	\hline    \end{tabular} \hspace{2em}
\begin{tabular}{llll}        \hline        \hline        $\Delta_{3}$ & $\Delta^{*}_{3}$ &&  coloring \\        \hline &&& \\          tetrahedron\ & {\em node} & {\footnotesize (4 links)} &$i_n=i_e$ 	  \\        triangle & {\em link} && $j_l=j_f$  \\        segment &  &&\\        point & &&\\         \hline        \hline    \end{tabular}\index{triangulation}\index{triangulation!dual}\index{4-simplex}\index{edge}\index{vertex}\index{face}\end{center}\label{Table2}   \end{table}

Finally, recall that the last expression can be interpreted as the quantum gravity boundary amplitude associated to the boundary state defined by the spin network $s$ \cite{book}. The individual spin foams $\sigma$ appearing in the sum can be interpreted as (discretized) spacetimes bounded by a 3-geometry determined by $s$. That is, (\ref{ssf}) can be interpreted as a concrete definition of the formal functional integral 
\begin{equation}
\Psi[q]  = \int_{\partial g = q} Dg \ e^{iS_{\rm GR}[g]}
\end{equation}
where $q$ is a 3-geometry and the integral of the exponent of the general relativity action is over the 4-geometries $g$ bounded by $q$.  Indeed, (\ref{ssf}) can also be derived from a discretization of a suitable formulation of this functional integral. We now turn to the physical interpretation of this boundary 3-geometry. 

\subsection{Relation with geometry}

In order to select a physically relevant boundary state $\Psi_{\mathbf q}[s]$, we need a geometrical interpretation of the boundary spin networks $s$. To this aim, recall that 
the spinfoam model can be obtained from a discretization of general relativity on a triangulated  spacetime.  The discretization can be obtained as follows. 

We associate an $R^ 4$ vector $e_{\rm s}^I$ to each segment $\rm s$ of the triangulation.  The relation with the gravitational field can be thought as follows.  Introduce 4d coordinates $x^\mu$ and represent the gravitational field by means of the one-form tetrad field $e^I(x) = e^I_\mu(x)dx^\mu$ (related to Einstein's metric by $g_{\mu\nu}(x)=e^I_\mu(x)e_I{}_\mu(x)$).   Assuming that the triangulation is fine enough for this field to be approximately constant on a tetrahedron, with constant value $e^I_\mu$, associate the 4d vector $e_{\rm s}^I=e^I_\mu
\Delta x^\mu_{\rm s}$ to the segment ${\rm s}$, where $\Delta x^\mu_{\rm s}$ is the coordinate difference between the initial and final extremes of ${\rm s}$. Next,  to each triangle $t$ of the triangulation, associate the bivector (that is, the object with two antisymmetric indices)
\begin{equation}B^{IJ}_{t} = e_{\rm s}^I e_{\rm s'}^J - e_{\rm s}^J e_{\rm s'}^I ,
\label{B}
\end{equation}
where $\rm s$ and $\rm s'$ are two sides of the triangle. (As far as orientation is kept consistent, the choice of the sides does not affect the definition of $B^{IJ}_{t}$). $B^{IJ}_{t} $ is a discretization of the Plebanski two-form $B^{IJ} = e^I\wedge e^J$.  The quantum theory is then formally obtained by choosing the quantities $B^{IJ}_{t}$ as basic variables, and identifying them with $SO(4)$ generators $J^{IJ}_{t}$ associated to each triangle of the triangulation, or, equivalently, to each face of the corresponding dual spinfoam. (For a compairaison with Regge calculus, see \cite{gionti}.) 

The geometry is then easily reconstructed using the $SO(4)$ Casimirs.  In particular, the peculiar form (\ref{B}) implies immediately that 
\begin{equation}
\epsilon_{IJKL}B^{IJ}_{t}B^{KL}_{t'}=0
\label{constr3}
\end{equation}
 any time $t=t'$ or $t$ and $t'$ share an edge. Accordingly, the pseudo--scalar Casimir $\tilde C=\epsilon_{IJKL}J^{IJ}_{t} J^{KL}_{t}=0$ is required to vanish. This determines the restriction to the simple representations, which are precisely the ones for which $\tilde C$ vanishes. 
 
 The scalar Casimir $C=\frac{1}{2}J^{IJ}_{t} J_{t}{}_{IJ}=\frac{1}{2}B^{IJ}_{t} B_{t}{}_{IJ}$, on the other hand, is easily recognized, using again (\ref{B}), as the square of the \emph{area} $A_{t}$ of the triangle $t$. Indeed, calling 
 $\alpha_{\rm ss'}$ the angle between $\rm s$ and $\rm s'$, we have:
 \begin{eqnarray}C&=&\frac{1}{2}B^{IJ}_{t} B_{t}{}_{IJ}= \frac{1}{2}(e_{\rm s}^I e_{\rm s'}^J - e_{\rm s}^J e_{\rm s'}^I)(e_{{\rm s}I} e_{{\rm s'}J} - e_{{\rm s}J} e_{{\rm s'}I})\nonumber \\
 &=& e_{\rm s}\cdot e_{\rm s}\  e_{\rm s'}\cdot e_{\rm s'} - (e_{\rm s}\cdot  e_{\rm s'})^2 = |e_{\rm s}|^2|\ e_{\rm s'}|^2\ (1-\cos^2\alpha_{\rm ss'})\nonumber \\
 &=& (|e_{\rm s}|\ |e_{\rm s'}|\ \sin\alpha_{\rm ss'})^ 2 = A_{t}^ 2. \end{eqnarray}
For simple representations, the value of $C$ is $j(j+1)$. The quantization of the geometrical area, with $j(j+1)$ eigenvalues is of course a key result of LQG, reappearing here in the context of the spinfoam models. It is the LQG result that assures us that we can interpret it as a physical discretization and not an artifact of the triangulation of spacetime. 

An explanation about units is needed. $B_t^{IJ}$ has units of a length square, hence $C$ has units  $[L]^ 4$. In the quantum theory, $B_t^{IJ}$ is identified with $J_t^{IJ}$ and $C$ has discrete eigenvalues. The identification requires evidently a scale to be fixed. This scale determines the Planck constant. A posteriori, we can simply reconstruct the correct scale by using again LQG, where the area eigenvalues are 
\begin{equation}
A_{j} = \frac{8\pi \gamma \hbar G }{c^3}  \sqrt{j(j+1)}
\label{size}
\end{equation}
where $\gamma$ is the Immirzi parameter, which we fix to unit below, together with the speed of light $c$.  This fixes the scale of the discretization (that is, it fixes the ``size" of the compact $SO(4)$ group in physical units). 

Next, consider two triangles sharing a side. Say the triangle $t$ has two sides: the segments ${\rm s}_1$ and ${\rm s}_2$ while the triangle $t'$ has two sides ${\rm s}_1$ and ${\rm s}_3$. Consider the action of the $SO(4)$ generators on the tensor product of the representation spaces associated to the two (faces dual to the two) triangles. This is given by the operators $J^{IJ}_{tt'}=J^{IJ}_{t}+J^{IJ}_{t'}$ (we omit the tensor with the identity operator in the notation)). Equation (\ref{constr3}), for $t\ne t'$ implies, with simple algebra, that  the pseudo--scalar Casimir $\tilde C_{tt'}=\epsilon_{IJKL}J^{IJ}_{tt'} J^{KL}_{tt'}$ vanishes as well. This implies that the tensor product of the two representations associated with the triangles $t$ and $t'$ is --again-- only allowed to contain simple representations.  Let $t$ and $t'$ be two of the four triangles of a given tetrahedron. In the dual picture, they correspond to two faces joining along an edge $e$ of the spinfoam.  Then $\tilde C_{tt'}$ is the pseudo--scalar Casimir of the virtual link that defines the intertwiner associated to this edge, under the pairing that pairs $t$ and $t'$.  The vanishing of  $\tilde C_{tt'}$ implies that this virtual link, as well, is labeled by a simple representation.  In the model we are considering all internal edges are labeled by the Barrett-Crane intertwiner, whose key property is precisely that it is a linear combination of virtual links with simple representations for any possible pairing of the four adjacent faces, thus  consistently with $\tilde C_{tt'}=0$. This is in fact \emph{la raison d'\^etre} of the Barrett--Crane intertwiner. 

Let us now consider the boundary $s$ of the spinfoam $\sigma$. A face $f$ that cuts the boundary, labelled by a simple representation $j_f$, defines a link $l$ of the boundary spin network $s$, equally colored with a representation $j_l=j_f$. As we have seen, the quantity $j_f(j_f+1)$ is to be interpreted as the area of the triangle dual to the face $f$. This triangle lies on the boundary and is cut by the link $l$. 

Notice that we have precisely the LQG result that the area of a triangle is determined by the spin associated to the link of the spin network that cuts it.  We can therefore identify in a natural way the boundary spin networks with the spin network states of canonical LQG.   Recall that in LQG a basis of states of the quantum geometry of a 3d surface is labelled by abstract spin networks $s$.  Since our aim here is not to fix the details of the physically correct quantum theory of gravity, but only to develop a general relativistic quantum formalism, we will do so in the following, disregarding some open issues raised by this identification (see below).  

The interpretation of the intertwiners at the boundaries is more delicate. Consider an edge $e$ of $\sigma$ that cuts the boundary at a node $n$ of $s$. The node $n$, or the edge $e$ are dual to a tetrahedron sitting on a boundary. Let  $t$ and $t'$ be two faces of this tetrahedron, and say, as above, that the triangle $t$ has two sides ${\rm s}_1$ and ${\rm s}_2$ while the triangle $t'$ has two sides ${\rm s}_1$ and ${\rm s}_3$.  Consider now the scalar Casimir $C_{tt'}=J^{IJ}_{tt'} J_{tt'\, IJ}$ on the tensor product of the representation spaces of the two triangles. Straightforward algebra shows that 
\begin{equation}
C_{tt'}=|C_t|+|C_{t'}|+2\ \vec n_t\cdot \vec n_{t'}.
\label{constr2}
\end{equation}
where $n^I_t=\epsilon^I{}_{JKL} B_t^{JK} t^L$ is and $t^L$ is the normalized vector  normal to $t$ and $t'$ (that is, to ${\rm s}_1,{\rm s}_2$ and ${\rm s}_3$).  Finally, $\vec n_t\cdot \vec n_{t'}=A_t A_{t'} \cos \alpha_{tt'}$, where $\alpha_{tt'}$ is the dihedral angle between $t$ and $t'$.     This provides the interpretation of the color of a virtual link  in the intertwiner associated to the node, in the corresponding decomposition: if the virtual 
link of this intertwiner is simple, with spin $j_{tt'}$, we have 
\begin{equation}
j_{tt'}(j_{tt'}+1)=A^2_t+A_{t'}^2+ A_t A_{t'} \cos \alpha_{tt'}.
\label{4}
\end{equation}
That is, the color of the virtual link is a quantum number determining the dihedral angle $\cos \alpha_{tt'}$ between the  triangles $t$ and $t'$; or, in the dual picture, the angle between the two corresponding links that join at $n$.   

Once more, this result is exactly the same in 3d LQG.  In this case, to each link is associated an $SU(2)$ generator $J^i, i = 1,2,3$, that can be identified with the $SU(2)$ valued two-form $E^i$ integrated on the dual triangle.  The color of the link is the quantum number of the $SU(2)$ Casimir 
${\mathbf C}=(J_t^i+J_{t'}^i)(J_{t\,i}+J_{t'\,i})$.  Expanding, we have $c=|J_t|^2+|J_{t'}|^2+2J_t^iJ_{t'\,i}$ or 
\begin{equation}
{\mathbf j_{tt'}}({\mathbf j_{tt'}}+1)=A^2_t+A_{t'}^2+ A_t A_{t'} \cos \alpha_{tt'}.
\label{3}
\end{equation}
where ${\mathbf j_{tt'}}$ is the quantum number labelling the eigenspaces of ${\mathbf C}$.  We are therefore lead to identify the intertwiner $i_{j_{tt'}}$ in the boundary spin network, with the intertwiner  $i_{\mathbf j_{tt'}}$ in the LQG spinnetwork states, since they represent the same physical quantity.  \label{appendice}

In fact, there is a key difference between (\ref{4}) and  (\ref{3}). In (\ref{4}), $j_{tt'}$ is the quantum number labelling a simple $SO(4)$ representation (recall $SO(4)$ irreducibles are labelled by pairs of spins, which are equal for simple representations); while in (\ref{3}),  ${\mathbf j_{tt'}}$ is the single spin labelling an $SU(2)$ representation. Some potential difficulties raised by this difference are discussed in Appendix B.   As argued in the Appendix, if we disregard these difficulties and we identify the intertwiner $i_{j_{tt'}}$ with the LQG intertwiner  $i_{\mathbf j_{tt'}}$, we obtain simply and consistently 
\begin{equation}\label{pippo} \langle i_{\mathbf j_{tt'}}  |i_{\rm BC}\rangle=(2{j_{tt'}} +1)= {\rm dim}(j_{tt'}). \end{equation}
The details of this interpretation do not play a role in this paper. We leave a more complete discussion of this issue open. 

This completes the geometrical interpretation of all quantities appearing in the spinfoam model. 

\subsection{Graviton operator}

The next ingredient we need is the graviton field operator.  This is the fluctuation of the metric operator over the flat metric.  At every point of the surface $\Sigma$ we chose a local frame in which the surface is locally stationary: three coordinates $x^a$ with $a=1,2,3$ coordinatize $\Sigma$ locally, and the metric is in the ``temporal" gauge: $g_{44}=1, g_{4a}=0$.  
To the first relevant order, we define $h^{ab}(\vec{x}) = g^{ab}(\vec{x}) - \delta^{ab}$. It is convenient to consider here the fluctuation of the \emph{densitized} metric operator 
\begin{equation}
\tilde h^{ab}(\vec{x}) = (\det{g}) g^{ab}(\vec{x}) - \delta^{ab}= E^{ai}(\vec{x})E^{bi}(\vec{x}) - \delta^{ab}.
\end{equation}
  In the linear theory, the propagators of the two agree because of the trace-free condition.  To determine its action, we can equally use the geometrical interpretation discussed above, or, directly, LQG.   We study the action of this operator on a boundary spin network state:  
\begin{equation}
E^{ai}(\vec{x})E^{bi}(\vec{x})|s\rangle. 
\end{equation}
Let us identify the point $\vec{x}$ with one of the nodes $n$ of the boundary spin network $s$.
Equivalently, with (the center of) one of the tetrahedra of the triangulation.  
Four links emerge from this vertex. 
Say these are $e_I, I=1,2,3,4$. They are dual to the faces of the corresponding tetrahedron. Let $n^I_a$ be the oriented normal to this face, defined as the vector product of two sides. 
Then $E(n)^{Ii}=E^{ai}(\vec{x})n_a^I$ can be identified with the action of the an $SU(2)$ generator $J^i$ on the edge $e_I$. We have then immediately that the diagonal terms define diagonal operators
\begin{equation}
E^{Ii}(n)E^I_{i}(n)|s\rangle = (8\pi \hbar G)^2\ j_{a} (j_{a} +1)|s\rangle 
\label{diagonal}
\end{equation}
where $j_a$ is the spin of the link in the direction $a$. 
The non--diagonal terms, that we do not consider in the following, are given in Appendix C.

\subsection{The boundary vacuum state}

As discussed in Section 2, the propagator will depend on a geometry $\mathbf q$ of the boundary surface $\Sigma$. Let us begin by choosing this 3d geometry.    Let $\mathbf q$ be isomorphic to the intrinsic and extrinsic geometry of the boundary $\Sigma_{\mathbf q}$ of a 4d (metric) ball in Euclidean $R^4$ with given radius, much larger than the Planck length.  We want to construct the state $\Psi_{\mathbf q}[s]$.
(On the vacuum states in LQG, see \cite{vuoto1,vuoto2,vuoto3,vuoto4,vuoto5,vuoto6}.)   Below we shall only need the value of  $\Psi_{\mathbf q}[s]$ for the spinnetworks $s=(\Gamma, j_{l}, i_n)$ defined on graphs $\Gamma$ which are dual to 3d triangulations $\Delta$.  We identify each such $\Delta$ with a fixed triangulation of $\Sigma_{\mathbf q}$. 

We assume here for simplicity that, for each graph,  $\Psi_{\mathbf q}[s]$ is given by a function of the spins of $s$ which is non-vanishing only on a single intertwiner on each node, which prjects on the $i_{\rm BC}$ intertwiner under (\ref{pippo}).  This will play no role in this paper, because, as we shall see, we compute only diagonal components of the propagator, which do not depend on the intertwiners. The precise role of the intertwiners, and other choices for the intertwiner dependence of the boundary state, will be discussed elsewhere.

The area $A_{l}$ of the triangle $t_{l}$ of $\Delta$, dual to the link $l$, determines background values ${j_{l}}^{\! \scriptscriptstyle (0)}$ of the spins $j_l$, via
 \begin{equation}
A_{l} = 8\pi\hbar G\ \sqrt{{j_{l}}^{\! \scriptscriptstyle (0)} ({j_{l}}^{\! \scriptscriptstyle (0)} +1)}.
\label{backgroundarea}
\end{equation}
We take these background values large with respect to the Planck length, and we will later consider only the dominant terms in $1/{j_{l}}^{\! \scriptscriptstyle (0)}$. 

We want a state $\Psi_{\mathbf q}[s]=\Psi_{\mathbf q}(\Gamma, {\mathbf j})$, where $ {\mathbf j}=\{j_l\}$, to be peaked on these background values.  The simplest possibility is to choose a Gaussian peaked on these values, for every graph $\Gamma$
\begin{equation}
\Psi_{\mathbf q}[s] = C_\Gamma\ 
 \exp\left\{-\frac{1}{2} \sum_{ll'}\alpha_{ll'}\ 
 \frac{j_{l} - {j_{l}}^{\! \scriptscriptstyle (0)}}{\left({j_{l}}^{\! \scriptscriptstyle (0)}\right)^{\frac k2}}\ 
 \frac{j_{l'} - {j_{l'}}^{\! \scriptscriptstyle (0)}}{\left({j_{l'}}^{\! \scriptscriptstyle (0)}\right)^{\frac k2}}
 + i \sum_{l}\Phi^{(0)}_{l} \, j_{l}\right\}
\label{vuotomenouno}
\end{equation} 
where $l$ runs on links of $s$, $\alpha_{ll'}$ is a given numerical matrix, $k\in(0,2)$ (see below),  and $C_\Gamma$ is a graph--dependent normalization factor for the gaussian. 

The phase factors in (\ref{vuotomenouno}) play an important role \cite{carlo}. As we know from elementary quantum mechanics, the phase of a semiclassical state determines where the state is peaked in the conjugate variables, here the variables conjugate to the spins $j_l$. Recall the form of the Regge action for one simplex, 
$S_{\rm Regge}=\sum_{l}\Phi_{l}(j_{l}) j_{l}$, where $\Phi_{l}(j_{l}) $ are the dihedral angles at the triangles\footnote{These are angles between the normals to the tetrahedra, and should not be confused with the angles between the normals to the faces, which are related to the intertwiners, as we discussed in Section 3.2.}, which are function of the areas themselves and recall that  
$ \partial S_{\rm Regge}/\partial j_{l}=\Phi_{l}$.
   It is then easy to see that these dihedral angles are precisely the variables conjugate to the spins. Notice that they code the extrinsic geometry of the boundary surface, and in GR the extrinsic curvature is indeed the variable conjugate to the 3-metric.   Thus, $\Phi^{(0)}_{l}$ are determined by the dihedral angles of the triangulation $\Delta$.  

Concerning the quadratic term in (\ref{vuotomenouno}) we have put the $(1/{j_{l}}^{\! \scriptscriptstyle (0)})^{k/2}$ factors in evidence because we want a semiclassical state for which the relative uncertainties of area and angle become small when all the areas are large, namely in the large distance limit in which all the spins ${j_{l}}^{\! \scriptscriptstyle (0)}$ are of the order of a large $j_L$.  That is, we demand that 
\begin{equation}
 \frac{\Delta A}{A} \to  0 \ \ \  {\rm and} \ \ \   \frac{\Delta \Phi}{\Phi} \to 0 , \ \ \ 
 {\rm when} \ \ \  {j_{l}}^{\! \scriptscriptstyle (0)}\sim j_L \to\infty. 
\label{demand}
\end{equation} 
Assuming that the matrix elements $\alpha_{(l)(l')}\sim \alpha$ do not scale with $j_L$, the fluctuations determined by the gaussian state (\ref{vuotomenouno}) are of the order
\begin{equation}
\Delta j \sim {\frac{j_L^{k/2}}{\sqrt\alpha}},  \hspace{2cm}  \Delta \Phi 
\sim {\frac{\sqrt\alpha}{j_L^{k/2}}} .
\end{equation} 
Therefore, since angles do not scale, 
\begin{equation}
\frac{\Delta A}{A} \sim \frac{\Delta j}{j} \sim {\frac{j_L^{k/2-1}}{\sqrt\alpha}},  \hspace{2cm}  \Delta \Phi \sim {\frac{\sqrt\alpha}{j_L^{k/2}}} .
\label{scale}
\end{equation} 
(\ref{demand}) and (\ref{scale}) restricts to $k\in(0,2)$.  From now on, we choose $k=1$. That is
\begin{equation}
\Psi_{\mathbf q}[s] = C_\Gamma\ 
 \exp\left\{-\frac{1}{2} \sum_{ll'}\alpha_{ll'}\ 
 \frac{j_{l} - {j_{l}}^{\! \scriptscriptstyle (0)}}{\sqrt{{j_{l}}^{\! \scriptscriptstyle (0)}}}\ 
 \frac{j_{l'} - {j_{l'}}^{\! \scriptscriptstyle (0)}}{\sqrt{ {j_{l'}}^{\! \scriptscriptstyle (0)}}}
 + i \sum_{l}\Phi^{(0)}_{l} \, j_{l}\right\}.
\label{vuoto}
\end{equation} 

The need for this dependence on the scale of the background of the covariance matrix of the vacuum state was been pointed out by one of us in the 3d context \cite{simone} and by John Baez in the 4d case, following numerical investigation by Dan Christensen and Greg Egan, that have shown that in the absence of this dependence the width of the gaussian is not sufficient for the approximation taken above to hold \cite{JDC}.

A strong constraint on the  graph--dependent constants $C_\Gamma$ and matrix $\alpha_{ll'}$ is given by the WdW condition  (\ref{normalization}), which requires the state to satisfy the dynamics.   The physical interpretation of the matrix $\alpha_{ll'}$ is rather obvious: it reflects the vacuum correlations, and is the analog of the covariance matrix in the exponent of the vacuum functional in the conventional Schr\"odinger representation of quantum field theory. The physical interpretation of the $C_\Gamma$ coefficients is less clear to us; it bears on the core of the diff-invariant physics of loop quantum gravity, and will be discussed elsewhere.   

\subsection{The 10$j$ symbol and its derivatives}

Baez, Christensen and Egan have performed in \cite{BCE} a detailed numerical analysis, which has lead them to conjecture that if we rescale all spins by a factor $\lambda$, then for large $\lambda$  the 10$j$ symbol can be expressed as a sum of two terms,
\begin{equation}B(j_{ij})= \sum_\sigma P(\sigma)\ \cos\left[S_{\rm Regge}(\sigma)+k\frac{\pi}{4}\right]+D(j_{ik}). 
\label{ansatz}\end{equation}
$P(\sigma)$ is a slowly varying factor, that grows as $\lambda^{-9/2}$ when scaling the spins by $\lambda$.  To understand this formula, consider a 4-simplex in $R^4$, with triangles $t_{ij}$ having areas $A_{ij}=\sqrt{j_{ij}(j_{ij}+1)}$. In general, there may be several distinct 4-simplices with triangles having these areas: let's label the distinct 4-simplices with a discrete label $\sigma$.   Each triangle $t_{ij}$ separates two boundary tetrahedra $\tau_i$ and $\tau_j$ of the 4-simplex. Each tetrahedron  $\tau_i$ defines a normal vector $n_i$, normalized and normal to all sides of the tetrahedron.  The angle $\Phi_{ij}$ between the normals $n_i$ and $n_j$ is the dihedral angle between the tetrahedra  $\tau_i$ and $\tau_j$. (The triangles $t_{ij}$ are in one-to-one correspondence with the links $l$ of the boundary spin network, hence the notation $\Phi_{ij}$ is consistent with the notation $\Phi_l$ used above.) For a fixed $\sigma$, we can compute the dihedral angles $\Phi_{ij}$ as a function $\Phi_{ij}(j_{ij})$, of the areas $A_{ij}$, hence of the 10 spins $j_{ij}$. The Regge action associated to the 10 spins is 
\begin{equation}S_{\rm Regge}(\sigma)= \sum_{ij} j_{ij}\ \Phi_{ij}(j_{ij}). \end{equation}
It is characterized by the fact that 
\begin{equation}\frac{\partial S_{\rm Regge}(\sigma)}{\partial j_{ij}}= \Phi_{ij}(j_{ij});\end{equation}
that is, the derivative with respect to the $j_{ij}$ in the angles does not contribute to the total derivative (this is the discrete analog to the fact that when we vary the Einstein-Hilbert action with respect to the metric, the metric variation of the Christoffel symbols  does not contribute.)

The form (\ref{ansatz}) for the 10$j$ symbol has later been confirmed by detailed analytical calculations by Barrett and Steele \cite{BS} and by Freidel and Louapre \cite{FL}.   As first noted in \cite{BarrettWilliams}, the first term in (\ref{ansatz}) is very good news for quantum gravity: it indicates that the 10$j$ symbols are indeed related to 4d general relativity.    

On the other hand, to understand the origin of the second term $D(j_{ik})$ in (\ref{ansatz}), recall that the 10$j$ symbols can be expressed in the form
\begin{equation}B(j_{ij})= \int_{(S^3)^5} dy_i \ 
\prod_{i<j} \frac{\sin((2j_{ij}+1)\Phi_{ij})}{\sin(\Phi_{ij})}. 
\label{intrep}\end{equation}
Here $\Phi_{ij}$ is the angle between the two unit vectors $y_i$ and $y_j$. The large $j$ behavior of this expression has been evaluated in  \cite{BS} and \cite{FL} using a stationary phase approximation. It turns out that the integration variables $y_i$ admit a very interesting geometrical  interpretation as the normals to the tetrahedra $y_i=n_i$.  The stationary points of the integral can be interpreted as different geometrical configurations.  Some stationary points are given by non degenerate tetrahedra.  These yield the Regge term in (\ref{ansatz}). But there are also contributions to the integral coming when two of the $y_i$ are parallel, or more in general when the linear span of the five $y_i$ is of dimension smaller than four.  These degenerate contributions yield the $D(j_{ik})$ term.  

The bad news is that this degenerate term strongly dominates for large $j$. This fact casts a thick shadow of doubt over hope that a Barrett-Crane spinfoam model could yield the correct general relativistic dynamics.  In fact, the discovery of the degenerate contributions is one of the sources of a recent decrease in interest in these models.
However, light comes back into the shadow, in consideration of the results of the previous section.   In fact, observe that what enters in the expression we have found in the previous section is not the 10$j$ symbol itself, but rather a second derivative of the 10$j$ symbol, because of the field insertions in (\ref{prodi}).  The fact that the degenerate term $D(j_{ij})$ dominates over the Regge term at large $j$ does not imply that its second derivative dominates as well.  

A hint that this hope may be correct comes from the following naive argument.  Degenerate contributions arise when the denominator in the integrand of (\ref{intrep}) vanishes. The formal second derivative with respect to $j_{ij}$, considered as a continuous variable, is 
\begin{equation}\frac{\partial^2 B(j_{ij})}{\partial j^2_{kl}}= - 4 \int_{(S^3)^5} dy_i \ 
  \Phi_{kl}  \prod_{i<j}\frac{\sin((2j_{ij}+1)\Phi_{ij})}{\sin(\Phi_{ij})}, 
\end{equation}
which decreases the order of the divergence in $\Phi_{ij}=0$.  A better version of this argument, due to Greg Egan \cite{Greg}, is the following. The diagonal term of the discrete second derivative
is  $B(j+1,...) - 2 B(j,...) + B(j-1,...)$.  Using the trigonometric identity:
\begin{equation}\sin((a+2)t) - 2 \sin(at) + \sin((a-2)t) = -4 \sin(t)^2 \sin(at)
\end{equation}
 we see that the 
effect of taking the diagonal discrete second derivative is to multiply the 
relevant kernel by $-4 \sin(\Phi)^2$.  This should then eliminate all the fully 
degenerate points from the integral. (The elimination of the fully degenerate points does not necessarily mean that the only remaining contribution is the Regge one, since, as shown in  \cite{BS}, there is a complex zoology of other degenerate contributions to the integral.  Numerical analysis to address some of this issues is being developed by Dan Christensen \cite{Dan}.)

\section{Order zero}

Let us begin by evaluating the general covariant 2-point function to order zero in $\lambda$. 
To this order 
\begin{equation}
W_0[s] = Z_0^{-1} \int \mathrm{D} \phi \, f_s(\phi) \, \mbox{e}^{-\int \phi^2}.
\end{equation}
The Wick expansion of this integral gives non--vanishing contributions for all $s$ with an even number of nodes.  Since there are no vertices, each of these contributions is simply given by products of face contributions, namely products of dimensions of representations. 
The 2-point function (\ref{simplypropaga1}) reads  
\begin{equation}
{ \mathbf G}^{abcd}_{\mathbf q}(\mathbf x, \mathbf y)=
\sum_s W_0[s] \ 
\hat h^{ab}(\mathbf x)
\hat h^{cd}(\mathbf y)
\Psi_{\mathbf q}[s].
\end{equation}
 Inserting (\ref{vuoto}), we have 
\begin{equation}
{ \mathbf G}^{abcd}_{\mathbf q}(\mathbf x, \mathbf y)=
\sum_s W_0[s] \ 
\hat h^{ab}(\mathbf x)
\hat h^{cd}(\mathbf y)\ 
C_\Gamma 
 \exp\left\{-\frac{1}{2} \sum_{ll'}\alpha_{ll'}\ 
 \frac{j_{l} - {j_{l}}^{\! \scriptscriptstyle (0)}}{\sqrt{{j_{l}}^{\! \scriptscriptstyle (0)}}}\ 
 \frac{j_{l'} - {j_{l'}}^{\! \scriptscriptstyle (0)}}{\sqrt{ {j_{l'}}^{\! \scriptscriptstyle (0)}}}
 + i \sum_{l}\Phi^{(0)}_{l} \, j_{l}\right\}.
 \end{equation}
We are interested in this expression for large $j^{\! \scriptscriptstyle (0)}_l$. In this regime,  
the gaussian effectively restricts the sum over (a large region of) spins of order  $j^{\! \scriptscriptstyle (0)}_l$. Over this region, the phase factor fluctuates widely, and suppresses the sum, unless it is compensated by a similar phase factor.  But  $W_0[s]$ contains only powers of $j_l$'s, and cannot provide this compensation.  Hence we do not expect a contribution of zero order to the sum.  The only exception can be the null spin network $s=\emptyset$, which gives 
$W[\emptyset]=1$ because of the normalization. Hence, to order zero
\begin{equation}
{ \mathbf G}^{abcd}_{\mathbf q}(\mathbf x, \mathbf y)=
W_0[\emptyset] \ 
\hat h^{ab}(\mathbf x)
\hat h^{cd}(\mathbf y)\ 
C_\emptyset.
 \end{equation}
But is reasonable to assume that the semiclassical boundary state on a macroscopic geometry $\mathbf q$ has vanishing component on $s=\emptyset$, whose interpretation is that of a quantum state without any volume. Hence we take $C_\emptyset=0$, and we conclude that the 2-point function has no zero order component in $\lambda$.

This result has a compelling geometrical interpretation.  The sum over spinfoams can be interpreted as a sum over 4-geometries.  The boundary state $\Psi_{\mathbf q}[s]$ describes a boundary geometry which has a nontrivial extrinsic curvature, described by the phase of the state. In the large distance limit, we expect semiclassical histories to dominate the path integral.  These must be close to a classical solution of the equations of motion, fitting the boundary data.   Because of the extrinsic curvature of the boundary data, it is necessary that the internal geometry has non vanishing 4-volume.  A round soccer ball must have volume inside.   But the four-volume of a spinfoam is given by its vertices, which are dual to the four-simplices of the triangulation.  Absence of vertices means absence of four-volume. It is therefore to be expected that the zero order contribution, which has no vertices, and therefore zero volume, is suppressed by the phases of the boundary state, representing the extrinsic curvature. 

Let us therefore go over to to the first order in $\lambda$.

\section{First order: the 4d nutshell}\label{firstorder}

Consider a spinfoam $\sigma$, dual to a single 4-simplex. Its the boundary spinnetwork has five nodes, connected by 10 links, forming the five-valent graph $\Gamma_5$. That is  
\begin{equation}
s = \hspace{-4em}\setlength{\unitlength}{0.0007in}
\begin{picture}(5198,1100)(5000,-4330)
\thicklines
\put(8101,-5161){\circle*{68}}
\put(8401,-3961){\circle*{68}}
\put(6601,-3961){\circle*{68}}
\put(7501,-3361){\circle*{68}}
\put(6901,-5161){\circle*{68}}
\put(8101,-5161){\line(1,4){300}}
\put(7501,-3361){\line( 3,-2){900}}
\put(6601,-3961){\line( 3, 2){900}}
\put(8101,-5161){\line(-1, 0){1200}}
\put(6901,-5161){\line(-1, 4){300}}
\put(7876,-4479){\line( 1,-3){225.800}}
\put(7089,-4351){\line(-6, 5){491.312}}
\put(7801,-3961){\line( 1, 0){600}}
\put(6901,-5161){\line( 1, 3){383.100}}
\put(7321,-3871){\line( 1, 3){173.100}}
\put(7501,-3354){\line( 1,-3){325.500}}
\put(7456,-4726){\line(-4,-3){581.920}}
\put(8394,-3961){\line(-5,-4){828.902}}
\put(7569,-4629){\line( 0,-1){  7}}
\put(6601,-3969){\line( 1, 0){1020}}
\put(8109,-5161){\line(-5, 4){867.073}}
\put(7456,-3264){\makebox(0,0)[lb]{\smash{${}_{i_1}$}}}
\put(8499,-3961){\makebox(0,0)[lb]{\smash{${}_{i_2}$}}}
\put(8229,-5326){\makebox(0,0)[lb]{\smash{${}_{i_3}$}}}
\put(6300,-3969){\makebox(0,0)[lb]{\smash{${}_{i_5}$}}}
\put(6714,-5349){\makebox(0,0)[lb]{\smash{${}_{i_4}$}}}
\put(7951,-3551){\makebox(0,0)[lb]{\smash{${}_{j_{12}}$}}}
\put(8349,-4644){\makebox(0,0)[lb]{\smash{${}_{j_{23}}$}}}
\put(7464,-5356){\makebox(0,0)[lb]{\smash{${}_{j_{34}}$}}}
\put(6504,-4696){\makebox(0,0)[lb]{\smash{${}_{j_{45}}$}}}
\put(6929,-3521){\makebox(0,0)[lb]{\smash{${}_{j_{51}}$}}}
\put(7569,-4261){\makebox(0,0)[lb]{\smash{${}_{j_{13}}$}}}
\put(7339,-4462){\makebox(0,0)[lb]{\smash{${}_{j_{35}}$}}}
\put(7266,-4284){\makebox(0,0)[lb]{\smash{${}_{j_{14}}$}}}
\put(7576,-4423){\makebox(0,0)[lb]{\smash{${}_{j_{24}}$}}}
\put(7397,-4110){\makebox(0,0)[lb]{\smash{${}_{j_{52}}$}}}
\end{picture}
\label{fsimpic}
\end{equation}
\vskip2cm\noindent
The boundary function $f_s(\phi)$ determined by this spin network is  
\begin{equation} 
f_s(\phi) =\sum_{\alpha_{nm}}
\phi^{\alpha_{12} \alpha_{13} \alpha_{14} i_{1}}
_{j_{12} j_{13} j_{14} j_{15}}\ 
\phi^{\alpha_{21} \alpha_{23} \alpha_{24} i_{2}}
_{j_{21} j_{23} j_{24} j_{25}}\ 
\phi^{\alpha_{31} \alpha_{32} \alpha_{34} i_{3}}
_{j_{31} j_{32} j_{34} j_{35}}\ 
\phi^{\alpha_{12} \alpha_{13} \alpha_{14} i_{4}}
_{j_{41} j_{42} j_{43} j_{45}}\ 
\phi^{\alpha_{51} \alpha_{52} \alpha_{53} i_5}
_{j_{51} j_{52} j_{53} j_{54}}\ 
\end{equation}
where $\alpha_{nm}\equiv\alpha_{mn}$. This can be compactly written as 
\begin{equation} 
f_s(\phi) = \sum_{\alpha_{nm}}\prod_{n=1,5} 
\phi^{\alpha_{nm}  i_n}_{j_{nm}}\ 
\end{equation}
This is a monomial of order five in the field, and is an observable in the group field theory. Its expectation value is given by (\ref{ev}). We consider the perturbative expansion (\ref{diliberto}).
At order $\lambda$, the only term remaining is
\begin{equation}
W[s] = 
 \frac{\lambda}{5!} \int \mathrm{D} \phi \, f_s(\phi) \, \left(\int \phi^5 \right) \, \mbox{e}^{-\int \phi^2}. 
\end{equation}
The Wick expansion of this integral gives one vertex $v$ and five propagators
\begin{equation}
W'[s] = 
 \frac{\lambda}{5!} 
 {\Big(\prod_{n=1,5}
 \mathcal{P}_{\alpha_{nm} i_n}^{j_{nm}}\,
 {}_{\alpha'_{nm}i_n'}^{j'_{nm}}
\Big)}
 \mathcal{V}_{\alpha'_{nm} i'_n}^{j'_{nm}},
 \end{equation}
where repeated representation indices are summed over. This expression still contains many terms due to the summation over the permutations in the propagators.   Recall that we can give the geometrical meaning of a face to each closed sequence of deltas in this expression. Each face contributes with a factor equal to the dimension of the representation. The dominant term for large representations is therefore the one with the largest number of surfaces. 
A short reflection will convince the reader that this is the term in which the surfaces correspond precisely at the faces of the dual of a four-simplex. That is, the dominant term of $W[s]$ to order $\lambda$ is 
\begin{equation}
W[s]  =  \frac{\lambda}{5!} \left( \prod_{n}   \langle i_n  |i_{\rm BC}\rangle   \right) 
\left(\prod_{n < m} {\rm dim}({j_{nm}})\right) \, 
                        \mathcal{B}(j^{\rm u}_{nm}).
\end{equation}    
Since we have chosen boundary states peaked on an intertwiner that projects on $i_{\rm BC}$, this reduces to 
\begin{equation}
W[s]  =  \frac{\lambda}{5!}
\left(\prod_{n < m} {\rm dim}({j_{nm}})\right) \, 
                        \mathcal{B}(j^{\rm u}_{nm}).
                                 \label{simply11}
\end{equation}    
This is the dominant term of the connected component of the amplitude for the boundary spin network considered, in the limit of large representations.  This is the expression we will use within equation  (\ref{simplypropaga1}).

The value of  $\Psi_{\mathbf q}[s]$ on the spin-networks $s=(\Gamma_5,  j_{nm})$ (here $n,m=1,...,5$) can be determined by triangulating $\Sigma_{\mathbf q}$ with the 3d triangulation formed by the boundary of a \emph{regular}  four--simplex of side $L$.
The area of the triangles is $A_L=\sqrt{3}L^2/4$.  Then  (\ref{backgroundarea}) implies that $j^{(0)}_{nm}=j_L$ where $8\pi \hbar G \sqrt{j_L(j_L+1)}=A_L$.  In the large $L$ limit we take $j_L=8\pi\hbar GA_L$.  The dihedral angles $\Phi_{nm}^{(0)}=\Phi$ of a regular tetrahedron are given by $\cos(\Phi) = -1/4$. Therefore (\ref{vuoto}) becomes 
\begin{equation}
\Psi_{\mathbf q}[s] = C_{5}\ 
 \exp\left\{-\frac{1}{2j_L} \sum_{(nm)(pq)}\alpha_{(nm)(pq)}\ 
(j_{nm} - j_L) (j_{pq} - j_L) + i \Phi \sum_{(n,m)}  j_{nm}\right\} .
\label{vuotofirst}
\end{equation} 
To respect the symmetry of the sphere, the covariance matrix $\alpha_{(nm)(pq)}$ of the gaussian can depend only on three numbers 
\begin{equation}
\alpha_{(nm)(pq)}= \alpha_1\ a_{(nm)(pq)}+ \alpha_2\ \delta_{(nm)(pq)}  +  \alpha_3\ b_{(nm)(pq)} 
\label{simmetry}
\end{equation}
where $\delta_{(nm)(pq)}=1$ if $(nm)=(pq)$,  $a_{(nm)(pq)}=1$ if just two indices are the same, 
and $b_{(nm)(pq)}=1$ if all four indices are different,  and in all other cases these quantities vanish.  We will use this notation, namely $\alpha_{(12)(13)}=\alpha_1, \ \alpha_{(12)(12)}=\alpha_2, \ \alpha_{(12)(34)}=\alpha_3 $ repeatedly. 

The component of the state (\ref{vuoto}) that matters at first order in $\lambda$ is thus completely determined up to the three numbers $\alpha_1,\alpha_2,\alpha_3$, and the constant $C_5$. This amounts to select a vacuum state which is a coherent state peaked both on the background values of the spins (the intrinsic geometry of the boundary surface), \emph{and} on the background values of the angles (the intrinsic geometry of the boundary surface). See \cite{nutshell} for a similar construction in 3d.

For clarity, let us stress that we are \emph{not} assuming that the boundary state has components \emph{only} on the five-valent graph considered.  What we are saying is that only this component of the boundary state enters the expansion to first order in $\lambda$ that we are considering. This has generated a certain confusion in public discussions of preliminary versions of this work. 

\subsection{First order graviton propagator}

We have now all the elements needed to compute the expression (\ref{simplypropaga1}).  Inserting 
(\ref{simply11}), (\ref{vuotofirst}) and (\ref{diagonal}) in  (\ref{simplypropaga1}) we obtain a completely well--defined expression for the propagator.   As a first step towards the analysis of the resulting expression, we  choose the points $\mathbf x$ and $\mathbf y$ to be two distinct nodes of the boundary spinnetwork.  Equivalently, these can be thought as points located, say, in the centers of the corresponding dual tetrahedra: in the theory, of course, position is not determined with better precision that the individual ``atoms  of spaces" described by the individual tetrahedra.  We consider the ten by ten matrix $\widetilde { \mathbf G}(L)$ formed by the ``diagonal" components of the propagator  
\begin{equation}
\widetilde { \mathbf G}(L)_{(ij)(kl)}\equiv  { \mathbf G}_{\mathbf q}^{abcd}(x,y)\ n^{\scriptscriptstyle(ij)}_a n^{\scriptscriptstyle(ij)}_b\  n^{\scriptscriptstyle(kl)}_c n^{\scriptscriptstyle(kl)}_d,
\end{equation}
where $n^{\scriptscriptstyle(ij)}_a$  is the normal to the triangle $t_{ij}$. Since all ten triangles have the same background area,  $|n^{\scriptscriptstyle(ij)}|=|n|=8\pi\hbar G j_{L}$ for large $j_L$,  we can write ${{ \mathbf G}(L)}  \equiv \widetilde { \mathbf G}(L)/{|n|^4}=\widetilde { \mathbf G}(L)/{(8\pi\hbar G j_{L})^4}$. By symmetry
\begin{equation}
{ \mathbf G}(L)_{(ij)(kl)}= { \mathbf G}_1(L)\ a_{(ij)(kl)}+ { \mathbf G}_2(L)\ \delta_{(ij)(kl)}  +  { \mathbf G}_3(L)\ b_{(ij)(kl)}. 
\label{simmetry2}
\end{equation}
This expression depends of course on $L$, which determines the distance between the points considered and the angles between the directions considered. 

Before computing this quantity in the background independent theory, let us compute it in conventional linearized quantum general relativity, for later comparison.  There are two ways of making the comparison.  One is to compare the propagator obtained in the background independent calculation with the linearized--theory propagator $G_{\rm linearized}^{abcd}(x,y)$, where $x$ and $y$ are in the \emph{center} of the corresponding tetrahedron.  The other is to compare it with the quantity obtained by \emph{integrating}  $G_{\rm linearized}^{abcd}(x,y)$ in $x$ and $y$, over the entire tetrahedron that they represent.   The difference is a numerical factor that is not relevant for us, and we choose here the first option.    In a flat background metric, two points in the center of adjacent tetrahedra, in a surface with the boundary geometry chosen, are at a distance ${|x-y|_q}=L/4$.   If the four indices $i,j,k,l$ are all distinct, it is easy to see that $n^{\scriptscriptstyle(ij)}$ and $n^{\scriptscriptstyle(kl)}$ are orthogonal;  then the propagator is easily computed to be 
\begin{equation}G_{(ij)(kl)}^{\rm linearized}(L) =i\frac {8\pi\hbar G}{4\pi^2}\frac{1}{|x-y|^2_q}=i\frac{32 \hbar G}{\pi L^2} 
\label{sivorrebbe1}
\end{equation}
On the other hand, the components $G^{\rm linearized}_{(ij)(ij)}$ and $G^{\rm linearized}_{(ij)(ik)}$ are vacuum expectation values at fixed ``time": the first is the fluctuation of the area square of a triangle, and the second is the vacuum correlation between the fluctuations of the area squares of two adjacent triangles in the same tetrahedron.  These are also proportional to $L^{-2}$. We can therefore write
\begin{equation}G^{\rm linearized}(L) =\frac{32 \hbar G}{\pi L^2}\ {\cal W} 
\label{sivorrebbe}
\end{equation}
where ${\cal W}$ is a numerical matrix, with the same symmetry structure as in (\ref{simmetry2}). 
For instance, ${\cal W}_{(12)(34)}=i$ as in (\ref{sivorrebbe1}), while the others projections are easily obtained from the linear theory. 

We now compute the matrix $ { \mathbf G}(L)$  in the full theory.  Since this is a diagonal term in the propagator, we can use (\ref{diagonal}) and (\ref{simplypropaga1}) reads
\begin{equation} { \mathbf G}(L)_{(ij)(kl)}=\frac{1}{8\pi\hbar G j_L^4}\sum_s W[s] \ 
((8\pi\hbar G)^2 j_{ij}(j_{ij}+1)-|n|^2)
((8\pi\hbar G)^2 j_{kl}(j_{kl}+1)-|\tilde n|^2)
\Psi_q[s].
\end{equation}
The terms $|n|^2$ come from the background $\delta^{ab}$ and are equal to the square of the area of the face, namely to $(8\pi\hbar G j_{L})^2$, for large $j$.   Inserting (\ref{simply2}) and (\ref{vuotofirst}) we have, to first order in $\lambda$
\begin{eqnarray}
{ \mathbf G}(L)_{(ij)(kl)}&=&
\frac{\lambda}{5!} \frac{1}{j_L^4} \sum_{j_{nm}} 
 \left(\prod_{n < m} {\rm dim}({j_{nm}})\right) \, 
(j_{ij}(j_{ij}+1)-j_{L}^2) \  ( j_{kl}(j_{kl}+1)-j_{L}^2) \ \ 
                         \nonumber\\   && 
                          \mathcal{B}(j_{nm})\ 
                         \ C_5 \ \exp \left\{
-\frac{1}{2j_L}\alpha_{(nm)(pq)}(j_{nm}- j_{L})(j_{pq}- j_{L})
+i\Phi \sum_{n<m} j_{nm}\right\},
\label{full}
\end{eqnarray} 
where we have used the  Einstein convention in the exponent. 
Since we have assumed that $j_{L}$ is large and the vacuum exponential peaks the sum around $j_{L}$, we can discard the $+1$ in the parenthesis.  We expand the summand in the fluctuations $\delta j_{ij}=(j_{ij}-j_{L})$, and keep only the lowest term, assuming that the gaussian suppress the higher terms.  
This gives 
\begin{eqnarray}{ \mathbf G}(L)_{(ij)(kl)}&=&
\frac{\lambda}{5!} \frac{4}{j_L^2}\sum_{j_{nm}} 
 \left(\prod_{n < m} {\rm dim}({j_{nm}})\right) \, 
\, \delta j_{nm}\, \delta_{pq}\ \ 
                         \nonumber\\   && 
                          \mathcal{B}(j_{nm})\ 
                         \ C_5 \ \exp \left\{
-\frac{1}{2j_L}\alpha_{(nm)(pq)}\, \delta j_{nm}\, \delta_{pq}
+i\Phi \sum_{n<m} j_{nm}\right\}. 
\label{full1}
\end{eqnarray} 
We assume that the ${\rm dim} j$ terms vary slowly over the range where the gaussian is peaked, and can be considered constant.  Let us absorb $C_5$ and these constants in a factor ${\cal N}_5$ .  
\begin{equation}{ \mathbf G}(L)_{(ij)(kl)}={\cal N}_5  
\frac{4}{j_L^2} \sum_{j_{nm}} 
\, \delta j_{nm}\, \delta_{pq}\ 
                          \mathcal{B}(j_{nm})\ 
                          \exp \left\{
-\frac{1}{2j_L}\alpha_{(nm)(pq)}\, \delta j_{nm}\, \delta_{pq}
+i\Phi \sum_{n<m} j_{nm}\right\}. 
\label{full2}
\end{equation} 
We change summation variable from the spins to the fluctuation of the spins 
\begin{equation}  { \mathbf G}(L)_{(ij)(kl)}={\cal N}_5 \frac{4}{j_L^2} 
\sum_{\delta j_{nm}} 
\delta j_{ij}\ \ \delta j_{kl} 
                     \    \mathcal{B}(j_L+\delta j_{nm})\ 
                         e^{-\frac{1}{2j_L}\alpha_{(nm)(pq)}\delta j_{nm}\delta j_{pq}
                                                  +i\Phi \sum_{nm}j_{nm}}. 
\label{somma}
\end{equation} 
 The sum can be approximated with a gaussian integral.
The rapidly oscillating term $\exp {i\Phi \sum_{nm}j_{nm}}$ tends to suppress the sum.  To evaluate it, we need the explicit form of $\mathcal{B}(j_L+\delta j_{nm})$ in the large $j$ regime, discussed above.   Since the sum (\ref{somma}) is peaked around $j_{nm}=j_L$, let us expand the 10$j$ symbol around this point. To second order around $j_{nm}=j_L$, the Regge action reads 
\begin{equation}S_{\rm Regge}(j_{nm}) = \Phi \sum_{nm} j_{nm} + \frac12 G_{(mn)(pq)} \delta j_{mn}\delta j_{pq}, \end{equation}
where, introducing the ``discrete derivative"
$\frac{\partial f(j)}{\partial j}\equiv \
f(j+1/2)-f(j)$, we have defined
\begin{equation}G_{(mn)(pq)}= \left.\frac{\partial\Phi_{mn}(j_{rs})}{\partial j_{pq}}\right|_{j_{rs}=j_L}. 
\label{angolispin}
\end{equation}
Thus, around  $j_{nm}=j_L$, (\ref{ansatz}) gives
\begin{equation}B(j_{nm})
= P_{\tau_R} \left[e^{i(\Phi\! \sum_{nm} j_{nm} + \frac12 G_{(nm)(pq)} \delta j_{nm}\delta j_{pq}+\frac{\pi}{4})}\!+e^{-i( \Phi\! \sum_{nm} j_{nm} + \frac12 G_{(nm)(pq)} \delta j_{nm}\delta j_{pq}+\frac{\pi}{4})}\right]\!+
D(j_{nm})
\end{equation}
where $\tau_R$ is the regular four simplex (for which $k_{\tau_R}=1$), which is the only non-degenerate four-simplex with these areas \cite{BCE}.  The key observation is now the fact that the rapidly oscillating $\exp \{i\Phi \sum_{nm}j_{nm}\}$ term in the second term of this expression cancels with the rapidly oscillating term in (\ref{somma}).  Therefore the second term of the last expression contributes in a non-negligible way to the sum 
(\ref{somma}). The first term is suppressed (by the rapidly oscillating factor $\exp {2i\Phi \sum_{nm}j_{nm}}$) and it is reasonable to expect that so is the degenerate term $D(j_{nm})$
because this corresponds to 4-simplices with different angles, and should
be dominated by different frequencies. This is the cancellation of the phases that 
we mentioned in Section 2.  Therefore  (\ref{somma}) becomes, keeping 
only the first term 
\begin{equation} { \mathbf G}(L)_{(ij)(kl)}={\cal N}'_5 \frac{4}{j_L^2} 
\sum_{\delta j_{nm}} 
\delta j_{ij}\ \ \delta j_{kl}  
\ e^{-\frac{i}{2} G_{(nm)(pq)} \delta j_{nm}\delta j_{pq}}
                         e^{-\frac{1}{2j_L}\alpha_{(nm)(pq)}\delta j_{nm}\delta j_{pq}},
\end{equation} 
where we have absorbed some constant factors in ${\cal N}'_5$. 
This factor, which contain the  constant $C_5$ of the state, is determined by the WdW condition (\ref{normalization}). This is simply given by the same expression without the field operator insertions, namely 
\begin{equation} 1= {\cal N}'_5
\sum_{\delta j_{nm}} 
\ e^{-\frac{i}{2} G_{(nm)(pq)} \delta j_{nm}\delta j_{pq}}
                         e^{-\frac{1}{2j_L}\alpha_{(nm)(pq)}\delta j_{nm}\delta j_{pq}}. 
\end{equation} 
Thus using the form (\ref{prodi}) for the two-point function, we have
\begin{equation} { \mathbf G}(L)_{(ij)(kl)}=\frac{4}{j_L^2} 
\frac{\sum_{\delta j_{nm}} 
\delta j_{ij}\ \ \delta j_{kl}  
\ e^{\left(-\frac{i}{2} G_{(nm)(pq)}-\frac{1}{2j_L}\alpha_{(nm)(pq)}\right)\delta j_{nm}\delta j_{pq}}}
{ \sum_{\delta j_{nm}} 
\ e^{\left(-\frac{i}{2} G_{(nm)(pq)}-\frac{1}{2j_L}\alpha_{(nm)(pq)}\right)\delta j_{nm}\delta j_{pq}}}.
\end{equation} 
This expression for the 2-point function has the same structure as equation (\ref{matrix2}).
Approximating the sum by a gaussian integral gives 
\begin{equation} { \mathbf G}(L)
= \frac{4}{j_L^2}   \left(j^{-1}_L\alpha + iG\right)^{-1}.
\label{eccola}
\end{equation} 

We only need to evaluate the derivatives (\ref{angolispin}) of the angles with respect to the spins.   4d geometry gives  (see the Appendix A)
\begin{equation}\label{kappa}
G_{(nm)(pq)}={8\pi\hbar G}\ \frac{\partial \Phi_{nm}}{\partial A_{pq}}= \frac{8\pi\hbar G}{\sqrt{5}L^2}\left(\frac72 a_{(nm)(pq)}-9 \delta_{(nm)(pq)}  -4 b_{(nm)(pq)}\right)\equiv \frac{8\pi\hbar G}{L^2}\ {\cal K}_{(nm)(pq)}.
\end{equation}
The ten by ten matrix ${\cal K}$ has purely numerical entries. From the relation between areas and spins, we have $j_L={\sqrt3 L^2}/(32\pi\hbar G)$. The $j_L$ factor that combines with the one in front of $\alpha$ in 
(\ref{eccola}) to give the crucial overall $1/L^2$ dependence of the propagator. Finally
(\ref{eccola}) reads
\begin{equation} { \mathbf G}(L) 
=\frac{32\pi\hbar G}{\sqrt{3}/4\; L^2}\left(\alpha+ i\sqrt{3}/4\; {\cal K}\right)^{-1}. 
\label{eccola2}
\end{equation} 
This is the value  (\ref{sivorrebbe1}-\ref{sivorrebbe}) of the propagator computed from the linearized theory, with the correct $1/|x-y|^2$ spacetime dependence. The three numerical coefficients of the matrix $\alpha$ are completely determined by $\alpha={{4\pi^2}}/{\sqrt{3}} \ {\cal W}^{-1} -i{\sqrt{3}}/{4}\ {\cal K}$. 

\section{Second order}

\begin{figure}[t]
  \begin{center}
  \includegraphics[height=6cm]{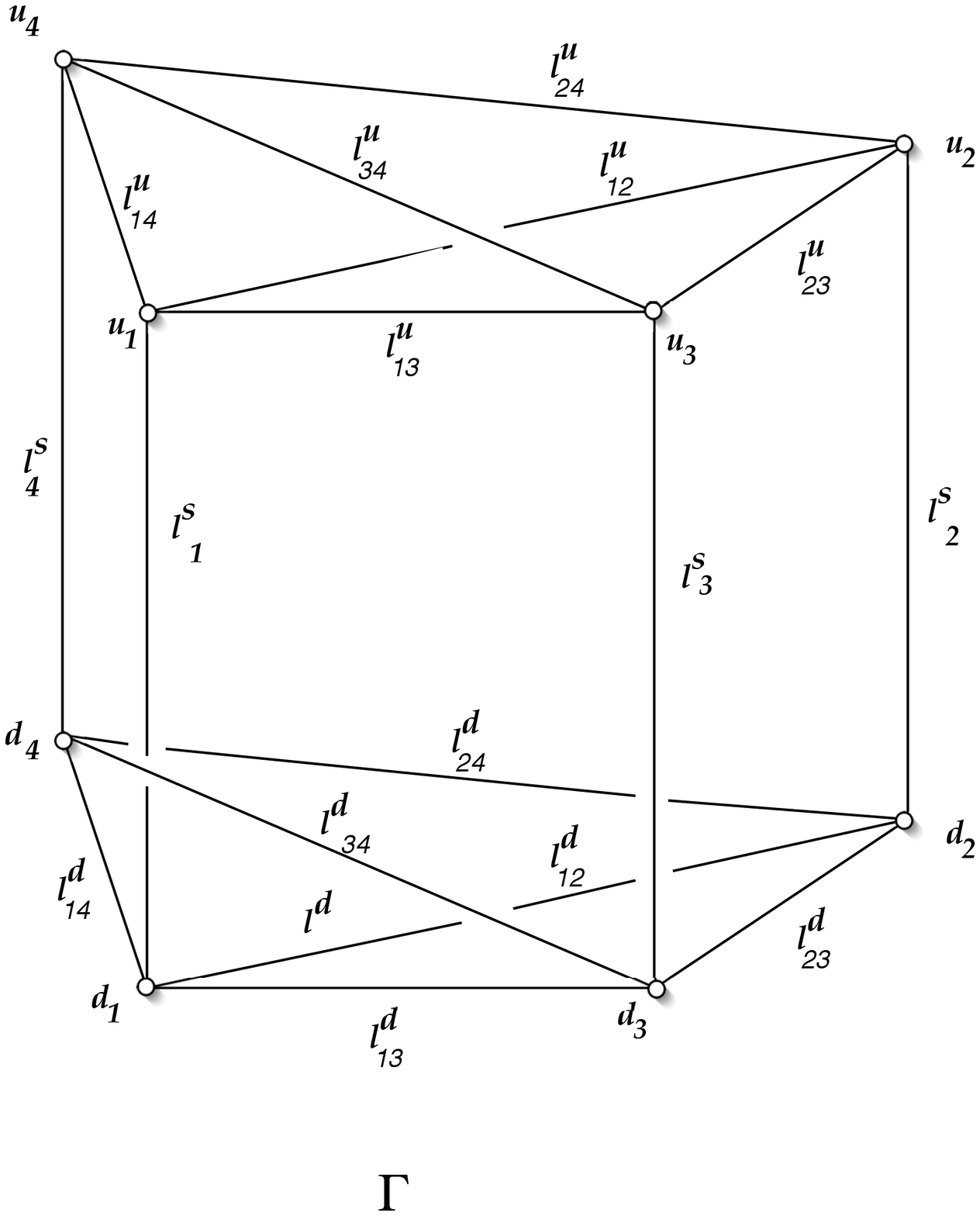}
\hspace{3em}
  \includegraphics[height=6cm]{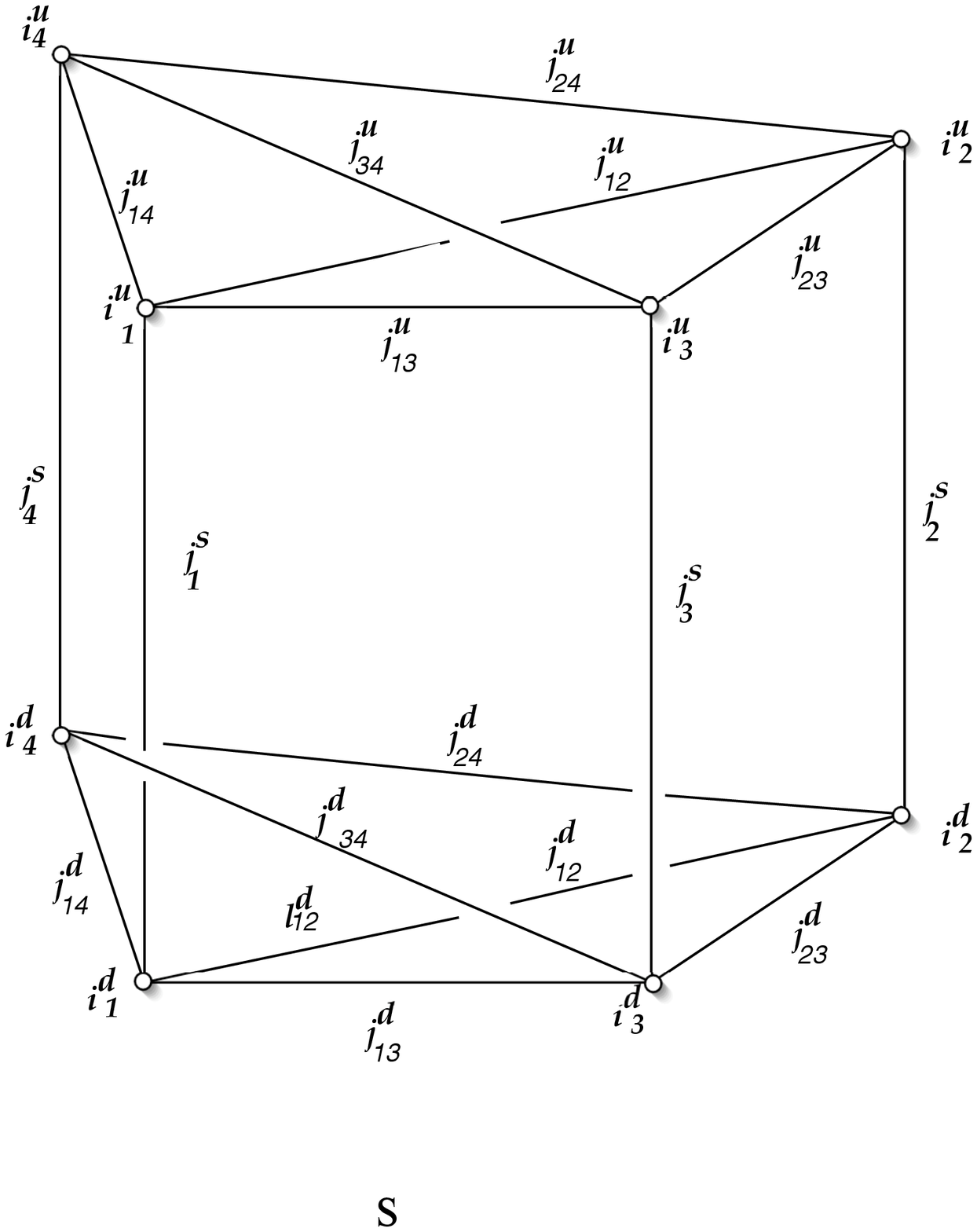}
  \end{center}
  \caption{\label{Boundary02} The boundary graph $\Gamma$ and the boundary spin network $s$.}
  \end{figure}

Next, we consider the second order term in the expansion in $\lambda$. We focus on one of the contributions, leaving the general case for further developments. 
The case we consider is determined by a boundary graph $\Gamma_8$ illustrated in Fig.\ref{Boundary02}.  It consists of two tetrahedral spin networks connected by four links.  Denote the nodes of the first tetrahedra $u_n$ and the nodes of the second one $d_n,\ n=1,...,4$ ($u$ for ``up", $d$ for ``down").  The links of $\Gamma$ are the six links $l^{\rm u}_{nm}$ with $n< m$ of the ``up" tetrahedron, the six links $l^{\rm d}_{nm}$ of the ``down" tetrahedron, and the four side links $l^{\rm s}_n$,  connecting $u_n$ with $d_n$.

\begin{figure}[b]
  \begin{center}
  \includegraphics[height=5cm]{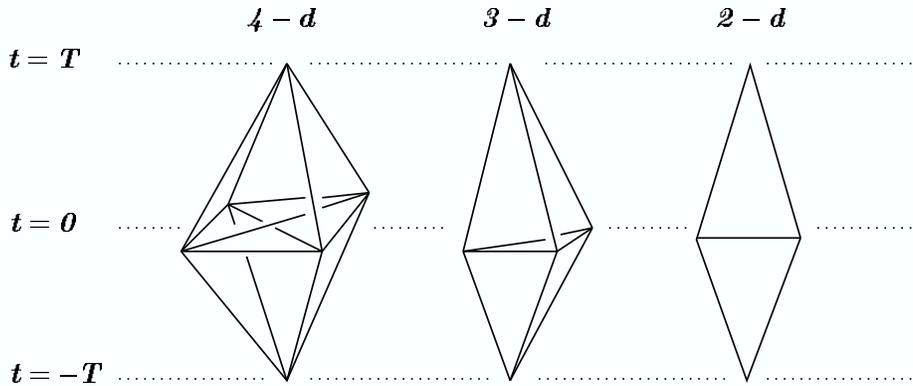}
  \end{center}
  \caption{\label{Analog} The spacetime triangulation $\Delta$ together with its 3d and 2d analogs.}
  \end{figure}

\begin{figure}[t]
  \begin{center}
  \includegraphics[height=5cm]{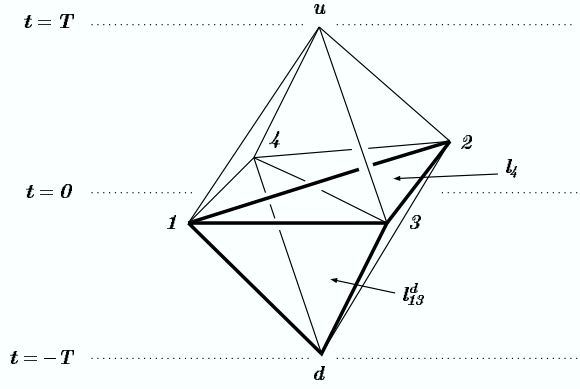}
  \end{center}
  \caption{\label{Dual0} The spacetime triangulation.}
  \end{figure}

In order to illustrate the geometrical meaning of this boundary graph, consider the four--dimensional triangulation $\Delta$ illustrated in the first panel of  Fig.\,\ref{Analog}, formed by two four-simplices (say the ``up" one $v^{\rm up}$ and the ``down" one  $v^{\rm down}$), joint by a tetrahedron $\mathcal{T}$. If we interpret the vertical axis as a ``time" axis, the triangulation $\Delta$ represents the world-history of a point $d$ opening up to a tetrahedron $\mathcal{T}$ and then recollapsing to a point $u$.    This is the 4d analog of the 3d and 2d cases illustrated in the other two panels of the figure.  In the 3d case, we have a point opening up to a triangle and then recollapsing;  in the 2d case, we have a point opening up to a segment and then recollapsing.  Label the four vertices of  $\mathcal{T}$ with an index $n=1,...,4$, call  $l^{\rm u}_{nm}$ (respectively $l^{\rm d}_{nm}$) the triangles formed by the vertices  of  $\mathcal{T}$ $n,m$ and $u$ (respectively $n,m$ and $d$), and $l_n$ the triangle opposite to the vertex $n$ of  $\mathcal{T}$. The four triangles $l^{\rm s}_n$ bound $\mathcal{T}$ (Fig. \ref{Dual0}).

\begin{figure}[t]
  \begin{center}
  \includegraphics[height=6cm]{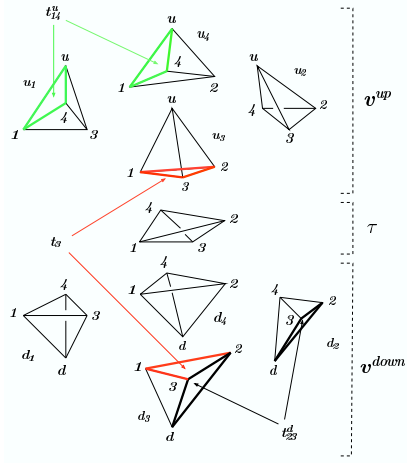}
  \end{center}
  \caption{\label{Decomposition} Decomposition}
  \end{figure}

Now, consider the \emph{boundary} of the triangulation $\Delta$. The eight boundary tetrahedra and the tetrahedron $\cal T$ are represented in Fig.\ref{Decomposition}. This is formed by four ``up" tetrahedra $u_n$ (the ones bounding $v^{\rm up}$, except for $\mathcal{T}$) and 
four ``down" tetrahedra $d_n$  (the ones bounding $v^{\rm down}$, except for $\mathcal{T}$). 
The two upper tetrahedra $u_n$ and $u_m$ are separated by the triangle $t^{\rm u}_{nm}$.
Similarly, the two lower tetrahedra are separated by a triangle $t^{\rm d}_{nm}$.  Finally, each upper tetrahedron is separated by a lower tetrahedron by a triangle $t_n$.

The \emph{dual} of the boundary of $\Delta$ is precisely $\Gamma$, defined above, and illustrated in  Fig.\ref{Boundary02}:  the four upper nodes of $\Gamma$ correspond to the four upper tetrahedra; the four lower nodes of $\Gamma$ correspond to the four lower tetrahedra. The six links joining the upper (lower) nodes correspond to the six upper (lower) vertical triangles $t^{\rm u}_{nm}$ ($t^{\rm d}_{nm}$); the four vertical links correspond to the four triangles $t_{n}$ bounding $\mathcal{T}$.  
In conclusion, $\Gamma$ is the dual triangulation of a 3d surface that can be viewed as the boundary of a spacetime region formed by a point expanding to a tetrahedron and then recollapsing to a point. 

Let us denote $j^{\rm u}_{nm},j^{\rm d}_{nm},j{\rm s}_{n}$ the twenty simple $SO(4)$ irreducible representations associated to the 20 links $l^u_{nm},l^d_{nm}, l^s_{n}$ of $\Gamma$, and $i_n^{\rm u}, i_n^{\rm d}$ eight intertwiners associated to the eight nodes $u_n$ and $d_n$.  The set $s=(\Gamma,  j^{\rm u}_{nm},j^{\rm d}_{nm},j^{\rm s}_{n}, i_n^{\rm u}, i_n^{\rm d})$ is the boundary spin network we consider in this section.

\begin{figure}[t]
  \begin{center}
  \includegraphics[height=5cm]{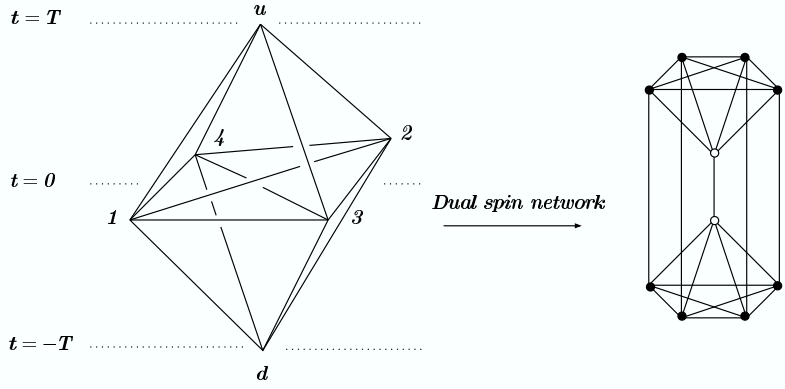}\hspace{3em}
  \end{center}
  \caption{\label{Dual} On the left, the triangulated spacetime. On, the right, its dual drawn around the Feynman graph.}
  \end{figure}

The boundary function $f_s(\phi)$ determined by this spin network is  
\begin{equation} 
f_s(\phi) = \sum_{\alpha_{nm}\beta_{nm}}\prod_{n=1,4} 
\phi^{\alpha_{nm}  i^{\rm u}_n}
_{j^{\rm u}_{nm}}\ 
\phi^{\beta_{nm}  i^{\rm d}_n}
_{j^{\rm d}_{nm}}
\end{equation}
This is a monomial of order eight in the field, and is an observable in the group field theory. Its expectation value is given by (\ref{ev}).  At order $\lambda^2$, this gives
\begin{equation}
W[s] = 
 \frac{\lambda^2}{2 (5!)^2} \int \mathrm{D} \phi \, f_s(\phi) \, \left(\int \phi^5 \right)^2 \, \mbox{e}^{-\int \phi^2}
\end{equation}
The Wick expansion of this integral gives two vertices, say $v^{\rm up}$ and $v^{\rm down}$ and nine propagators. Namely the Feynman graph
 \begin{center}
  \includegraphics[height=2cm]{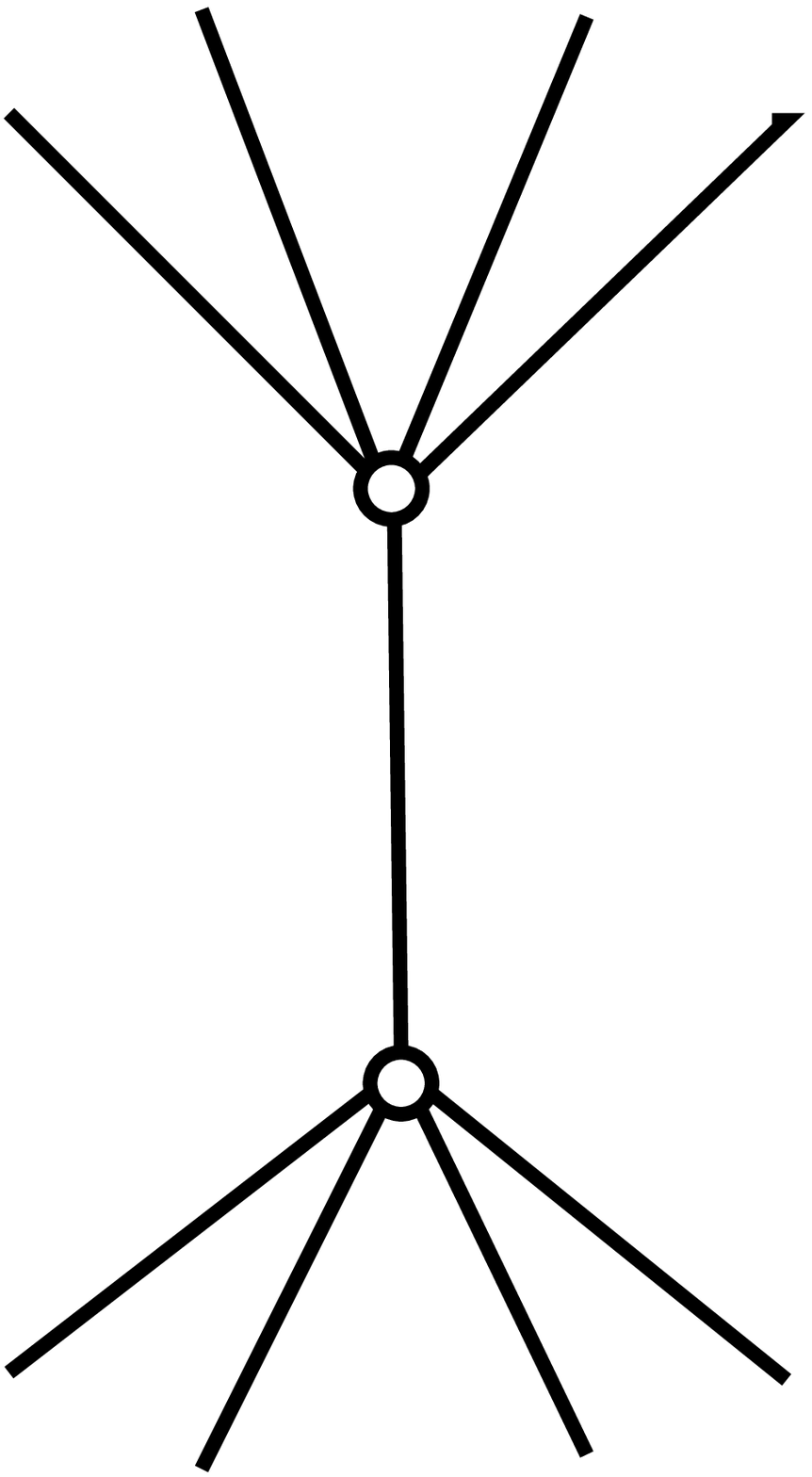}
  \end{center}
Let us focus on the particular term of the Wick expansion obtained contracting the two vertices with one propagator, and contracting the other four legs of $v^{\rm up}$ (respectively $v^{\rm down}$)  with the four ``up" nodes $u_n$ (respectively, with the four ``down" nodes $d_n$) of $s$, as represented in Fig.\ref{Dual}. This is the term
\begin{equation}
W'[s] = 
 \frac{\lambda^2}{2 (5!)^2} 
 {\Big(\prod_{n=1,4}
 \mathcal{P}_{\alpha_{nm}\alpha_n i^{\rm u}_n}^{j^{\rm u}_{nm}j_n}\,
 {}_{\alpha'_{nm}i_n'}^{j'_{nm}}
\Big)}
 \mathcal{V}_{\alpha'_{nm}\gamma_m i_n'i''}^{j'_{nm}j''_{m}}
 \mathcal{P}_{\gamma_{m} i''}^{j''_{m}}\,{}_{\delta_{m}i'''}^{j'''_{m}}
 \mathcal{V}_{\delta_m i''' \beta'_{nm} i_n'}^{j'''_{m} j''_{nm}}
 {\Big(\prod_{n=1,4}
 \mathcal{P}_{\beta_{nm}\beta_n i^{\rm d}_n}^{j^{\rm d}_{nm}j_n}\,{}_{\beta'_{nm}i_n'}^{j''_{nm}}
\Big)},
\end{equation}
where repeated representation indices are summed over. This expression still contains many terms due to the summation over the permutations in the propagators.  To find the dominant contribution, recall that each closed sequence of deltas in this expression is interpreted as face. Each face contributes with a factor equal to the dimension of the representation. The dominant  term for large representations is therefore the one with the largest number of surfaces. 
A short reflection will convince the reader that this is the term in which the faces is the spinfoam which is dual to the triangulation described above.  See Figure 3. That is, the dominant term is 
\begin{equation}
W[s]  =  \frac{\lambda^2}{2 (5!)^2} \left( \prod_{n}   \langle i^{\rm u}_n  |i_{\rm BC}\rangle   \langle i^{\rm d}_n  |i_{\rm BC}\rangle   
                 {\rm dim}({j_n})\right) \left(\prod_{n < m} {\rm dim}({j^{\rm u}_{nm}}) \,
                    {\rm dim}({j^{\rm d}_{nm}})\right) \, 
                        \mathcal{B}(j^{\rm u}_{nm},j_n),
\mathcal{B}(j^{\rm d}_{nm},j_n).  
         \label{simply3}
          \end{equation}    
If we chose the intertwiners as above, this reduces to           
\begin{equation}
W[s]  =  \frac{\lambda^2}{2 (5!)^2} \left( \prod_{n}  
                 {\rm dim}({j_n})\right) \left(\prod_{n < m} {\rm dim}({j^{\rm u}_{nm}}) \,
                    {\rm dim}({j^{\rm d}_{nm}})\right) \, 
                        \mathcal{B}(j^{\rm u}_{nm},j_n),
\mathcal{B}(j^{\rm d}_{nm},j_n).  
         \label{simply2}
          \end{equation}    

This is the dominant term of the connected component of the amplitude for the boundary spin network considered, in the limit of large representations.  This is the expression we will use within equation  (\ref{simplypropaga1}).

The boundary spin network is determined by the quantum numbers $(j_{nm}^{\rm u,d}, j^{\rm s}_n, i^{\rm u,d}_n)$.   The spins $j_{nm}^{\rm u,d}$ and $j^{\rm s}_{n}$ are the quantum numbers of the areas of the triangles $t_{nm}^{\rm u,d}$ and $t^{\rm s}_{n}$ respectively.   The intertwiners $i^{\rm u,d}_n$ are the quantum numbers of the dihedral angles between the triangles $t_n$ and $t_{n,(n+1)}^{\rm u, d}$.  

\subsection{Vacuum boundary state}

We now have to extend the choice of the vacuum boundary state $\Psi_{\mathbf q}[s]$ considered in the previous section to the larger spin networks considered here.   For simplicity, let us again fix the intertwiners to be Barrett-Crane intertwiners, as above.  We need therefore to select the function 
\begin{equation}\Psi_{\mathbf q}[s]=\Psi_{\mathbf q}( j^{\rm u,d}_{nm},j^{\rm s}_{n}). \end{equation}
We want this function to be peaked on background values $( j^{\rm u,d(0)}_{nm}, j^{\rm s(0)}_{n})$ of its arguments representing a given background geometry $\mathbf q$. Let this background geometry be the boundary triangulation by it being the boundary of the region of $R^4$ formed by the two pyramidal four-simplices having for basis a regular tetrahedron $\mathcal T$ with side of length $L'$, and height $T$.  

For this, let us consider a Euclidean 4d space with cartesian coordinates $t,x,y,z$. Consider a regular tetrahedron $\mathcal T$ in the $t=0$ surface, having side length $L$ and center on the origin of $E$.   Consider the two points $u=(T,0,0,0)$ and $d=(-T,0,0,0)$. The two 4-simplices bounded by $u$ (respect $d$) and $\mathcal T$, taken together, define a compact 4d region. Call $\Sigma$ the boundary of this region. $\Sigma$ is a  triangulated  3d \emph{metric} surface.  Let $\mathbf q$ be the geometry of $\Sigma$.  This is the classical metric on which  we want the state $\Psi_{\mathbf q}[s]$ to be peaked.
The barycenters of the eight tetrahedra sit on a sphere of radius $R'=\sqrt{(\frac T4)^2+(\frac34 \frac{1}{2\sqrt{6}}L')^2}$  and the barycenters of the two tetrahedra sharing the same face of $\mathcal T$ are at a distance $D'=\frac T2$ from each other (see Appendix A). These two barycenters are the point $\mathbf x$ and $\mathbf y$ where the 2-pont function is computed. In order to match the second order calculation with the first order one, we want the geometrical relations between these two points to be the same as in the first order calculation.  That is, we want 
$R'$ and $D'$ to be the same as in the first order calculation.
For this, we have to choose (see Appendix)
\begin{equation}
T=\frac L2, \hspace{4em} L'=\sqrt{\frac 52}\ L.
\label{fix}
\end{equation}
However, instead of fixing $L'$ and $T$ in this manner, let us keep them independent for the moment.   Intuitively, $L'$ represents the ``spatial" extension and $2T$ the ``temporal" duration of the spacetime region considered.

The area of the triangles of the triangulation is easily computed from elementary geometry.  The area of each of the triangles $t^{\rm s}_n$ is 
\begin{equation}
\label{AL}A_L=\frac{\sqrt{3}}{4}{L'}^2,\end{equation}
while the area of the each of the triangles $t_{nm}^{\rm u}$ and $t_{nm}^{\rm d}$
is  \begin{equation}
A_{LT} = 4 \, \frac{{L'}}{2} \sqrt{\frac{{L'}^2}{8} + T^2} + \frac{\sqrt{3}}{2} {L'}^2,
\label{ALT}\label{areageo}
\end{equation}
Equation (\ref{size}) gives us immediately the background values of the spins of $s$. Since we are interested in the large $j$ regime, we have
\begin{eqnarray}
j^{\rm s(0)}_n&=&   \frac{\sqrt{3}}{32\pi \hbar G }{L'}^2 \equiv j_L,   
\label{jL}
\\
 j^{\rm u(0)}_{nm}=j^{\rm d(0)}_{nm}&=&  \frac{1}{8\pi  \hbar G }   \left(2{L'} \sqrt{\frac{{L'}^2}{8} + T^2} + \frac{\sqrt{3}}{2} {L'}^2\right) \equiv j_{TL}.
\label{jTL}
\end{eqnarray}
We do not give here the explicit value of the background dihedral angles $\Phi_l^{\scriptscriptstyle (0)}$. This value can be obtained by elementary geometry and plays no role in the following.

\begin{figure}
  \begin{center}
  \includegraphics[height=4cm]{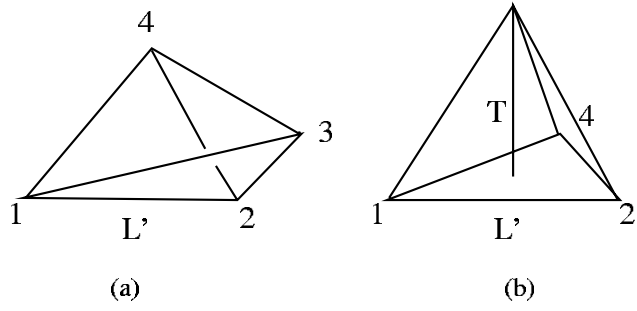}
  \end{center}
  \caption{\label{TetraLunghezze} Central tetrahedron $(a)$. Lateral tetrahedron $(b)$.}
  \end{figure}
  
The spin network $s^{(0)}=(\Gamma_8, j^{\rm u(0)}_{nm},j^{\rm d(0)}_{nm},j^{\rm s(0)}_{n})$ represents the discretization of the geometry $\mathbf q$ on the triangulation $\Delta$ of the metric surface $\Sigma$ (where we are disregarding the intertwiners). It will be our reference background spin network.   As above, we choose a Gaussian peaked on $s^{(0)}$. Writing all spins in a single vector $j_l=(j_{nm}^{\rm u},j_{nm}^{\rm d}, j^{\rm s}_{n})$, we have   
\begin{equation}
\Psi_{\mathbf q}[s] = 
C_8 \ e^{
- \alpha_{ll'}{(j_l- j_l^{\scriptscriptstyle (0)})(j_{l'}- j_{l'}^{\scriptscriptstyle (0)})}
+i \Phi_l^{\scriptscriptstyle (0)}j_{l}}. 
\label{vuoto2}
\end{equation} 
where we have used $\tilde\alpha_{ll'}=\frac{\alpha_{ll'}}{\sqrt{j^{ (0)}_{l}j^{ (0)}_{l'}}}$, and the Einstein convention.

\subsection{Second order graviton propagator}

Inserting (\ref{simply2}), (\ref{vuoto2}) and (\ref{diagonal}) in (\ref{simplypropaga1}) we obtain a well--defined expression for the propagator at  second order in $\lambda$.  Choose the points $\mathbf x$ and $\mathbf y$ to be nodes $u_1$ and $d_1$ of the boundary spin network.  Equivalently, these can be thought as coordinates located in the corresponding dual tetrahedra $u_1$ and $d_1$.   We thus consider the projection  
\begin{equation}\widetilde {\mathbf G}(L',T)={\mathbf G}_{\mathbf q}^{abcd}(\mathbf x, \mathbf y)\ n_an_b\  \tilde n_c \tilde n_d,\end{equation}
of the propagator. Notice that  $|n| =|\tilde n|$ and that $n_a$ and $\tilde n_a$ are orthogonal in the background metric.  We write ${\mathbf G}({L'},T)=\widetilde {\mathbf G}({L'},T)/|n|^4$ as above. 

Let's evaluate this quantity in linearized quantum gravity.  In the background metric considered, the two points have the same ``spatial" coordinates and a ``temporal" separation $T/2$ (the distance between the center of the two tetrahedra).  Hence
\begin{equation}{\mathbf G}^{\rm linearized}({L',T}) =\frac{8 \hbar G}{\pi T^2} 
\label{sivorrebbe3}
\end{equation}
which is the same as (\ref{sivorrebbe}) if (\ref{fix}) holds.

We now compute this quantity in the full theory. 
Using (\ref{diagonal}),  (\ref{simplypropaga1}) reads
\begin{equation} {\mathbf G}({L'},T)=\frac{1}{8\pi\hbar G j_{TL}^4}\sum_s W[s] \ 
((8\pi\hbar G)^2 j^{\rm u}_{12}(j^{\rm u}_{12}+1)-|n|^2)
((8\pi\hbar G)^2 j^{\rm u}_{13}(j^{\rm d}_{13}+1)-|\tilde n|^2)
\Psi_{\mathbf q}[s].
\end{equation} 
Inserting (\ref{simply2}) and (\ref{vuoto2}) we have 
\begin{eqnarray}
{\mathbf G}(L',T)&=&\frac{1}{j_{TL}^4} 
\frac{\lambda^2}{2 (5!)^2} \sum_{j_{l}} 
\prod_{l} {\rm dim}({j_{l}}) \, 
(j^{\rm u}_{12}(j^{\rm u}_{12}+1)-(j_{T{L}})^2) \  ( j^{\rm d}_{13}(j^{\rm d}_{13}+1)-(j_{TL})^2)
                         \nonumber\\   && 
 \ \ 
                          \mathcal{B}(j^{\rm u}_{nm},j^{\rm s}_n)\ 
\mathcal{B}(j^{\rm d}_{nm},j^{\rm s}_n)\ \ 
C_8  \ \exp \left\{
-\tilde\alpha_{ll'}(j_l- j_l^{\scriptscriptstyle (0)})(j_{l'}- j_{l'}^{\scriptscriptstyle (0)})
+i \Phi_l^{\scriptscriptstyle (0)}j_{l}\right\}. 
\end{eqnarray} 
Steps analogous to the first order case lead to
\begin{eqnarray} 
&& \hspace{-1cm} 
{\mathbf G}({L'},T) = 
\frac{4\lambda^2{\cal N}_8 }{2(5!)^2  j_{TL}^2} 
\sum_{\delta j_{l}} 
\delta j^{\rm u}_{12}\ \ \delta j^{\rm d}_{13} 
                     \   \mathcal{B}(j^{(0)}+\delta j_{l})\ 
\mathcal{B}(j^{(0)}+\delta j_{l})\ e^{
-\tilde\alpha_{ll'}(j_l- j_l^{\scriptscriptstyle (0)})(j_{l'}- j_{l'}^{\scriptscriptstyle (0)})+ i \,
 \Phi^{(0)}_{l} j_l },
\label{paperino}
\end{eqnarray} 
where a certain number of constants, including $C_8$ have been absorbed into ${\cal N}_8$.

We now use the asymptotic expression (\ref{ansatz}) for each $ \mathcal{B}$, obtaining
a total action given by the sum of Regge actions for the two 4-simplicies,
$S_{Regge}^{u}+S_{Regge}^{d}$. In this resulting action we can treat all spins as independent variables, and expand it around $j_{TL}$ and $j_L$
\begin{equation}
S_{Regge}(j_{nm}^{\rm u},j_{nm}^{\rm d}, j^{\rm s}_n) =
\tilde \phi^{(0)}_{n m} j_{n m}^{\rm u} +
 \tilde \phi^{(0)}_n j^{\rm s}_n + 
\tilde \phi^{(0)}_{n m} j_{n m}^{\rm d} +
 \tilde \phi^{(0)}_n j^{\rm s}_n 
+ \frac{1}{2} G_{ll'} \, \delta j_l \delta j_{l'}, 
\end{equation}
where $\tilde \phi^{(0)}_n$ and $\tilde \phi^{(0) }_{n m}$ are the dihedral angles of flat 4-simplices with the given boundary geometry; and the second order ``discrete derivatives" are 
\begin{equation}
G_{ll'} = \left(\frac{\delta^2 S_{Regge}}{\delta j_{l} 
\, \delta j _{l'}}\right)_{j_l  = j^{(0)}_l} .
\end{equation}
This matrix can be computed from elementary geometry, as we did in Section 5.1 (see also the Appendix). Being a derivative of an angle with respect of an area, for dimensional reasons, $G_{ll'}$ should scale as the inverse of $\sqrt{j^{(0)}_l j^{(0)}_{l'}}$. It is therefore convenient to define the scaled quantity 
\begin{equation}
\Gamma_{ll'} = \frac{G_{ll'}}{\sqrt{j^{(0)}_l j^{(0)}_{l'}}}.
\end{equation}

We assume, as before, that the only surviving term in the sum (\ref{paperino})
is the one in which the exponential in the cosine matches the phase of the boundary state.  Notice that this happens because the linear terms in the expansion of the Regge action sum up, giving the dihedral angle of the boundary of the 4d region, which is precisely the sum of the dihedral angles of the two 4-simplices at the faces of $\cal T$. That is, $\tilde\phi^{(0)}_{nm} = \Phi^{(0)}_{nm}$, but   $2 \tilde \phi^{(0)}_{n}=\Phi^{(0)}_{n}$.  Thus, we can rewrite  (\ref{paperino}) as 
\begin{equation}
{\mathbf G}({L'},T)=\frac{4 \lambda^2 {\cal N}_8 }{2 (5!)^2 \, j_{TL}^2} \, 
 \sum_{\delta j_{l}} 
\delta j^{\rm u}_{12} \,  \delta j^{\rm d}_{13}  \, 
P_{\tau}^2 \, e^{-i\left(S_{Regge}^u + S_{Regge}^d 
+ k_{\tau} \frac{\pi}{2}\right)} \mbox{e}^{
-\tilde\alpha_{ll'}(j_l- j_l^{\scriptscriptstyle (0)})(j_{l'}- j_{l'}^{\scriptscriptstyle (0)})
+ i  \Phi^{(0)} j_l}
\end{equation} 
The first order term of the expansion of the Regge action cancels the phases in the state, leaving
\begin{eqnarray}
{\mathbf G}({L'},T)&=&\frac{4 \lambda^2 {\cal N}_8 P_{\tau}^2\mbox{e}^{i k_{\tau} \frac{\pi}{2}}}{2 (5!)^2 j_{TL}^2}
 \sum_{\delta j_{l}} 
\delta j^{\rm u}_{12} \,  \delta j^{\rm d}_{13} \,
\mbox{e}^{-\tilde\alpha_{ll'}(j_l- j_l^{\scriptscriptstyle (0)})(j_{l'}- j_{l'}^{\scriptscriptstyle (0)})
- \frac{i}{2} G_{ll'} \, \delta j_l \, \delta j_{l'}} .
\end{eqnarray} 
The normalization factor is determined by (\ref{normalizzazione}).  Let us first provisionally assume  that $C_8$ is the only non-vanishing  $C_\Gamma$ coefficient in (\ref{prodi}). Then 
\begin{eqnarray}
&& {\cal N}_8^{-1} =\frac{\lambda^2 P_{\tau}^2 \, e^{i k_{\tau} \frac{\pi}{2}}}{2 (5!)^2}
 \sum_{\delta j_{l}} 
 \mbox{e}^{
-\tilde \alpha_{ll'}
\delta j_l \delta j_{l'}
} \ \mbox{e}^{-\frac{i}{2} G_{ll'} \, \delta j_l \, \delta j_{l'}}.  \end{eqnarray} 
We introduce the matrix 
\begin{eqnarray}
\tilde\mathcal{A}_{ll'} = 2\tilde\alpha_{ll'} +i  G_{ll'} = \sqrt{j^{(0)}_l j^{(0)}_{l'}}\ (2\alpha +i  \Gamma)_{ll'}  
 = \sqrt{j^{(0)}_l j^{(0)}_{l'}}\ \mathcal{A}_{ll'} 
\label{matrix}
\end{eqnarray}
and the vector $\delta{\bf j} =(\delta j_l)= (\delta j^{\rm u}, \delta j^{\rm d}, \delta j^{\rm s})$. So
\begin{eqnarray}
 {\mathbf G}(L',T)=\frac{4{\cal N}_8 P_{\tau}^2 \lambda^2 \mbox{e}^{i k_{\tau} \frac{\pi}{2}}}{2 (5!)^2 j_{TL}^2}
 \sum_{\delta j_{l}} 
\delta j^{\rm u}_{12} \,  \delta j^{\rm d}_{13} \,
\mbox{e}^{- \frac{1}{2} \delta {\bf j}^T \tilde\mathcal{A} \, \delta{\bf j}},
 \end{eqnarray} 
\begin{eqnarray}
{\cal N}_8^{-1} =\frac{\lambda^2 P_{\tau}^2 \, e^{i k_{\tau} \frac{\pi}{2}}}{2 (5!)^2}
 \sum_{\delta j_l} 
 \mbox{e}^{- \frac{1}{2} \delta {\bf j}^T \tilde\mathcal{A} \, \delta{\bf j}}.
\end{eqnarray}
And approximating the sum with gaussian integrals gives 
\begin{eqnarray}
{\mathbf G}(L', T) = \frac{16 \pi}{j_{TL}^2}
 \big(\tilde\mathcal{A}\big)^{-1}_{j_{12}^u, j_{13}^d}= 
\frac{16 \pi}{j_{TL}}
 \big(\mathcal{A}\big)^{-1}_{j_{12}^u, j_{13}^d}
\end{eqnarray}
Under the condition (\ref{fix}), ${\mathbf G}(L)={\mathbf G}(L'(L),T(L))$ is proportional to $1/j_L$, as in the first order calculation
Thus we recover the expected  $\frac{1}{L^2} $ behavior of the linearized theory.  The result is not proportional to $\lambda$. 

What if $C_8$ is not the only non-vanishing $C_\Gamma$ constant?  To illustrate this more general case, let us assume that only $C_5$ and $C_8$, namely the two terms considered at order one and two are non vanishing.  Then, calling  ${\mathbf G}_5(L)$ and  ${\mathbf G}_8(L)$ the 2-point function computed with the sole $C_5$ and $C_8$ terms, respectively, we have that 
the 2-point function has the structure
\begin{equation}
{\mathbf G}(L) = \lambda C_5 a_5 {\mathbf G}_5(L) + \lambda^2 C_8 a_8 {\mathbf G}_8(L),
\end{equation}
for appropriate constants $a_5$ and $a_8$; while the WdW condition (\ref{normalization}) reads 
\begin{equation}
1= \lambda C_5 a_5 + \lambda^2 C_8 a_8. 
\end{equation}
Posing ${\mathbf G}_8(L)={\mathbf G}_5(L)+\mathbf\Delta_8(L)$, and 
$k=(C_8 a_8)/ (C_5 a_5)$
we can write, to order $\lambda$
\begin{eqnarray}
{\mathbf G}(L) &=& \frac{\lambda C_5 a_5 {\mathbf G}_5(L) + \lambda^2 C_8 a_8 {\mathbf G}_8(L)}
{ \lambda C_5 a_5 + \lambda^2 C_8 a_8}
\nonumber \\ &=&  \frac{ {\mathbf G}_5(L) + \lambda k {\mathbf G}_8(L)}{1 + \lambda k}
\nonumber \\ &=&  {\mathbf G}_5(L) + \lambda k {\mathbf \Delta}_8(L).
\end{eqnarray}
Therefore, if the second order term gives the same result as the first order term, the overall normalization makes it irrelevant.  If, on the other hand, it gives a different result, then this appears as a correction of order $\lambda$ to the 2-point function. 

\section{Conclusion}

We have computed a first and a second order term in the expansion in $\lambda$ of the diagonal components of the graviton propagator, in a large distance regime, starting from a background--independent formulation of quantum gravity.  The result has the $1/|x-y|^2$ dependence on the distance expected from the linearized quantum theory, the expected dependence on the physical constants, and the numerical proportionality constants can be fixed as a condition on the semiclassical boundary state. The main tool we have used is the definition of general covariant $n$-point function, given in (\ref{eccolo4}). 

Many issues remain open.  Among these are: the calculation of non-diagonal terms in the propagator \cite{matrice}; the precise physical interpretation of the two expansion parameters, $\lambda$ and $1/j_L$ \cite{freidel};  the physical interpretation of the numerous subdominant terms, and their relation with the relativistic and the quantum corrections to the Newton law  \cite{SE};  a clarification of the implementation of the $SO(4)\to SO(3)$ gauge fixing at the boundary \cite{sergei} and of the overall Lorentz invariance of the formalism \cite{Rovelli:2002vp}; a 
precise formulation of the physical meaning of the observability of the boundary geometry (which is a partial observable \cite{book,HJ1,HJ,partial}); the possibility of computing  general covariant $n$-point function for models without expansion parameter $\lambda$; the full exploration of the dynamical WdW condition (\ref{normalization}) on the boundary state.  These issues be discussed elsewhere. 

In our opinion, the interest of the calculation is not so much in the final agreement with the linearized expression or in the details of the model used, but rather in the fact that it shows how some low-energy quantities with a transparent physical meaning can be computed, starting from the abstract context of a background independent formalism.    The specific choices of ingredients used for the calculation is therefore of lesser interest, in our opinion, that the display of the feasibility of calculations of this sort.

\vskip.5cm
\centerline{-----------------------------}\vskip.5cm

We thank Davide Mamone for some important suggestions, 
and Dan Christensen for his interest and some important
inputs based on the numerical analysis of the problem that he is developing. 
We also thank Emanuele Alesci, Federico Mattei and Massimo Testa for 
several discussions on this topic. 

\newpage
\appendix

\noindent{\bf \Large Appendix}

\section{Regular simplices}
\subsection{Elementary 4d geometry}

We collect here some simple geometrical formulas used in the text. An equilateral triangle of side $L$ has area $A=\frac{\sqrt{3}}{4}L^2$. An equilateral tetrahedron of side $L$ has volume $V_3=\frac{1}{6\sqrt{2}}L^3$ and height $h=\sqrt{\frac23}L$. The barycenter of the tetrahedron is at a distance $d=\frac h4=\frac{1}{2\sqrt{6}}L$ from a face. 

A regular 4-simplex of side $L$ has 4-volume $V_4=\frac{\sqrt{5}}{96}L^4$. The dihedral angles $\Theta$ of the 4-simplex, defined as the angles between the outward normals to the tetrahedra, satisfy $\cos{\Theta}=-1/4$. The center of two tetrahedra are at a distance $D=\frac L4$ from one another and at a distance $R=\frac{1}{2\sqrt{10}}$ from the center of the 4-simplex.

Consider an equilateral tetrahedron $\cal T$ in the $t=0$ plane, with center at the origin and side $L'$.  Consider the two points $u$ and $d$ with coordinates $t=T$ and, respectively, $t=-T$, and spatial coordinates in the origin.  Fix one of the faces of $\cal T$, and consider the tetrahedron defined by this face and the point $u$. The height of this tetrahedron joins the point $u$ with the center of the face, which is on $t=0$ and at a distance $d=\frac{1}{2\sqrt{6}}L'$ from the origin. The barycenter of a tetrahedron cuts this height in the proportion $3:1$, hence has coordinates $t=\frac  T4$ and $|x|=\frac34 \frac{1}{2\sqrt{6}}L'$. It is therefore at a distance $R'=\sqrt{(\frac T4)^2+(\frac34 \frac{1}{2\sqrt{6}}L')^2}$ from the origin and at a distance $D'=\frac T2$ from the barycenter of the tetrahedron defined by the face and the point $d$. 

If we demand that $D'=D$ and $R'=R$, we obtain $T=\frac L2$ and $L'=\sqrt{\frac 52}L$.

\subsection{The variation of the dihedral-angle with the area  in a regular 4d simplex}
Consider a 4-simplex in Euclidean 4d space.  Label its vertices with an index $i=1,...,5.$  Let $t_i$ be the tetrahedron opposite to the vertex $i$ and $\hat n_i$ a normalized vector normal to this tetrahedron and pointing outside the 4-simplex.  Then the (external) dihedral angle $\Phi_{ij}$ is given by 
\begin{equation}
              \cos{\Phi_{ij}}= \hat n_i \cdot \hat n_i. 
\label{cos}
\end{equation}
We can fix the geometry of the tetrahedron by giving the ten lengths $L_{ij}$ of the ten sides $(i,j)$. Alternatively, we can fix the geometry of the tetrahedron by giving the ten areas $A_{ij}$, where  $A_{ij}$ is the area of the triangle opposite to the two vertices $i$ and $j$.  If we do so, the dihedral angles are functions  $\Phi_{ij}(A_{kl})$ of the areas. We now consider the ten by ten jacobian matrix
$$
            K_{(ij)(lm)}(A_{kl}) = \frac{\partial \Phi_{ij}}{\partial A_{lm}}.
$$ 
In particular, using the notation of \Ref{kappa}, let 
\begin{equation}
            \f1{L^2}{\cal K}_{(ij)(lm)} = \left.\frac{\partial \Phi_{ij}}{\partial A_{lm}}\right|_{A_{lm}=A_L}. 
\label{fine}
\end{equation}
be the value of this jacobian on a regular tetrahedron of equal sides $L_{ij}=L$ and therefore areas $A_{ij}=A_L=\frac{\sqrt{3}}{4}L^2$.  
Here we present a simple way of computing it, which can be easily generalized to more complicated situations. 

We start from the relation 
\begin{equation}
            \sin{\Phi_{ij}}T_iT_j = \frac43 V A_{ij},
\label{relazione}
\end{equation}
where $V$ is the 4-volume of the simplex and $T_i$ is the volume of the tetrahedron $t_i$. This relation can be derived as follows. 
Pick one vertex, say $i=5$ as the origin of the coordinates, and choose coordinates such that the four sides originating from this vertex define the coordinate axes and the unit length. Let $g_{ab}$ be the Euclidean metric in these coordinates.  The triangle $(3,4)$, bounded by the vertices (5,1,2) is characterized by the bivector $\vec v_{12}^{\mu\nu}=1/2 (\delta_1^\mu \delta^\nu_2-\delta_2^\mu \delta^\nu_1)$ and its area is 
$$
A_{34}=|\vec v_{12}|=1/2\sqrt{g_{11}g_{22}-(g_{12})^2}. 
$$
The normal to the tetrahedron $t_a$ is given by $n^a_\mu=1/6\delta^a_\mu$ and the volume of the tetrahedron is its length, namely $T_a^2=|n^a|^2=1/6^2\ g^{aa}$.  
Equation (\ref{cos}) gives
\begin{equation}
              \cos{\Phi_{ab}}= \frac{n_a \cdot n_b}{|n_a|\; | n_b|}= \frac{g^{ab}}{\sqrt{g^{aa}g^{bb}}}. 
\end{equation}
Hence 
$$
            (\sin{\Phi_{34}}T_3T_4)^2 =(1-\cos^2{\Phi_{34}})(T_3T_4)^2 = 6^4 (g^{33}g^{44}-(g^{34})^2)=6^4
             \det g (g_{11}g_{22}-(g_{12})^2). 
$$ 
which is (\ref{relazione}), because the 4-volume is 1/24 $\sqrt{\det g}$. (The overall 3/4 factor can be easily checked by taking a rectangular 4-simplex of side 1.)

Varying (\ref{relazione}), we have 
\begin{equation}
            \cos{\Phi_{ij}}d\Phi_{ij} = \frac43 \frac{V A_{ij}}{T_iT_j}\left(\frac{dV}{V}+\frac{dA_{ij}}{A_{ij}}-\frac{dT_i}{T_i}+\frac{dT_j}{T_j}\right)=  \sin{\Phi_{ij}} \left(\frac{dV}{V}+\frac{dA_{ij}}{A_{ij}}-\frac{dT_i}{T_i}+\frac{dT_j}{T_j}\right).
\end{equation}
We now consider this relation for a regular 4-simplex. In this case,  $\cos{\Phi_{ij}}=-1/4$, therefore
\begin{equation}
            d\Phi_{ij} = -\sqrt{15} \left(\frac{dV}{V}+\frac{dA_{ij}}{A_{ij}}-\frac{dT_i}{T_i}+\frac{dT_j}{T_j}\right).
\label{vary}
\end{equation}

First, let us vary only one single length $L_{ij}$, keeping the other nine lengths fixed.  If we vary {\em all} lengths, then by dimensional analysis
$$
\frac{dV}{V}=4\frac{dL}{L},\ \ \ \ 
\frac{dA_{ij}}{A_{ij}}=2\frac{dL}{L},\ \ \ \ 
\frac{dT_i}{T_i}=3\frac{dL}{L}.
$$
If, instead, we vary {\em a single} side, we have by symmetry
\begin{equation}
\frac{dV}{V}=\frac{4}{10}\frac{dL}{L},\ \ \ \ 
\frac{dA_{ij}}{A_{ij}}=\frac{2}{3}\frac{dL}{L},\ \ \ \ 
\frac{dT_i}{T_i}=\frac{3}{6}\frac{dL}{L}.
\label{var2}
\end{equation}
because there are ten sides in a 4 simplex, three in a triangle and six in a tetrahedron. If we now vary the length of one single side in the 4 simplex, then the variation of the angle $\Phi_{ij}$ will be different according to the relative position of the angle and the side, that is, according to whether or not the side belongs or does not belong to the triangle $A_{ij}$ or the tetrahedra $T_j$. Explicitly, (\ref{vary}) and (\ref{var2}) give
\begin{equation}
            d\Phi_{ij} = -\sqrt{15} \left(\frac{4}{10} + \frac{2}{3}\epsilon_{ij}
          -\frac{3}{6}(\epsilon_i+\epsilon_j)\right)\frac{dL}{L}.
\end{equation}
where $\epsilon_{ij}=1$ if the side varied is in the triangle $A_{ij}$ and zero otherwise and
where $\epsilon_{i}=1$ if the side varied is in the tetrahedron $T_{i}$ and zero otherwise. 

With respect to a given angle $\Phi_{ij}$, there are three kinds of sides. One side $(i,j)$, six sides $(i,k)$ with $k\ne j$, and three sides $(k,l)$ with $k,l\ne i,j$.   In the first case, the side does not belong neither to $A_{ij}$ nor to $T_{i}$ or  $T_{j}$.  Hence
\begin{equation}
            \frac{\partial\Phi_{ij}}{\partial L_{ij}} = -\sqrt{15} \left(\frac{4}{10}\right)\frac{dL}{L}.
            \end{equation}
In the second case,  the sides belongs to one tetrahedron, but not to the triangle, hence
\begin{equation}
            \frac{\partial\Phi_{ij}}{\partial L_{ik}} = -\sqrt{15} \left(\frac{4}{10}-\frac{3}{6}\right)\frac{1}{L}.\end{equation}
In the third case, the side is in the triangle and both tetrahedra
\begin{equation}
            \frac{\partial\Phi_{ij}}{\partial L_{kl}} = -\sqrt{15} \left(\frac{4}{10}+\frac23-2\frac{3}{6}\right)\frac{1}{L}.
\end{equation}
In these equation and below, different indices are assumed to take different values and there is no sum over repeated indices. Summarizing
\begin{equation}
            \frac{\partial\Phi_{ij}}{\partial L_{ij}} =-2\sqrt{\frac35} \frac{1}{L}, \ \ \ \ \ \ 
            \frac{\partial\Phi_{ij}}{\partial L_{ik}} =  \frac{1}{2} \sqrt{\frac35} \frac{1}{L}, \ \ \ \ \ \ \ 
            \frac{\partial\Phi_{ij}}{\partial L_{kl}} = -\frac{1}{3} \sqrt{\frac35} \frac{1}{L}. 
\label{uno}
\end{equation}.

The next step is to compute the variation of the sides when one area varies, at constant value of the other areas.  By symmetry, if only the area of one triangle changes, then the variation of the three sides belonging to this triangle will be equal to themselves. Let $\delta L_{\rm triangle}$ be such variation. Similarly, the variation of the six sides that have only one vertex in common be triangle will be equal to themselves. Let $\delta L_{\rm side}$ be such variation. Finally, let 
$\delta L_{\rm opposite}$ be the variation of the single side opposite to the triangle. 

In an equilateral triangle the three variation of the the sides contribute equally to the variation of the area.  Consider a triangle adjacent to the triangle that varies. Its variation is proportional to $\delta L_{\rm triangle}+2\delta L_{\rm side}$ and vanishes.  Hence $\delta L_{\rm side}=-1/2 \delta L_{\rm triangle}$.  The same argument applied to the triangles with no side in common to the varying triangle gives $\delta L_{\rm opposite}=-2\delta L_{\rm side}=\delta L_{\rm triangle}$. Therefore we need only to compute $\delta L_{\rm triangle}$. This is easy, because for an equilateral triangle 
$\delta A/A=2\delta L /L$.   Hence $\delta L =L/2 \delta A /A=\frac{2}{\sqrt 3} \delta A/L^2$. Summarizing:
\begin{equation}
\frac{\partial L_{ij}}{\partial A_{kl}} = 2\ \frac{1}{\sqrt{3}}\ \frac{1}{L}, \ \ \ \ \ \ 
\frac{\partial L_{ij}}{\partial A_{ik}} =-\frac12\ \frac{1}{\sqrt{3}}\  \frac{1}{L}, \ \ \ \ \ \ 
\frac{\partial L_{ij}}{\partial A_{ij}} = 2\ \frac{1}{\sqrt{3}}\  \frac{1}{L}.
\label{duedue}
\end{equation}
The last step of the calculation is to combine (\ref{uno}) and (\ref{due})
\begin{equation}
\frac{\partial \Phi_{ij}}{\partial A_{kl}} = \sum_{nm} \frac{\partial \Phi_{ij}}{\partial L_{nm}} 
\frac{\partial L_{nm}}{\partial A_{kl}}, 
\end{equation}
where care should be taken, for each term of the sum to use the appropriate value depending on whether or not there are repeated indices. This is straightforward and gives
\begin{equation}
\frac{\partial \Phi_{ij}}{\partial A_{kl}} = -4\  \frac{1}{\sqrt{5}L^2}, \ \ \ \ \ \ 
\frac{\partial  \Phi_{ij}}{\partial A_{ik}} =\frac72\ \frac{1}{\sqrt{5} L^2}, \ \ \ \ \ \ 
\frac{\partial  \Phi_{ij}}{\partial A_{ij}} = -{9}\ \frac{1}{\sqrt{5}L^2}.
\label{due}
\end{equation}

In conclusion, we can write (\ref{fine}) as 
\begin{equation}
\f1{L^2}{\cal K}_{(ij)(kl)}= \frac{1}{\sqrt{5}L^2}\left(\frac72 a_{ij}^{kl}-4 \delta_{ij}^{kl}  -9 b_{ij}^{kl}\right)
\end{equation}
where $\delta_{ij}^{kl}=1$ if the two couples of indices are the same, 
$a_{ij}^{kl}=1$ if two and only two indices are the same, 
and $b_{ij}^{kl}=1$ if all four indices are different,  and in all other cases these quantities vanish.

\section{Boundary intertwiners}

At the end of Section \ref{appendice}, we have mentioned the difference between the space of the simple $SO(4)$ intertwiners and the space of the $SU(2)$ intertwiners.   The difference shows up in the linear structure of the states. Suppose we decide to pair the four faces of the tetrahedron differently, and to represent the intertwiner in terms of the virtual link $k=j_{tt''}$ for a different pairing.  In both cases, the linear properties of the space of the intertwiner allows to express a virtual link as a linear combination of virtual links of a different pairing, but the linear structure is different.  In fact, in the $SU(2)$ case, the recouping theorem gives (see for instance (A.65) of \cite{book}, also for the notation.)
\vskip.1cm
\begin{eqnarray}
i_{\mathbf j} = 
\begin{array}{c}\setlength{\unitlength}{1 pt}\begin{picture}(50,40)          \put( 0,0){$\mathbf j_1$}\put( 0,30){$\mathbf j_2$}          \put(45,0){$\mathbf j_4$}\put(45,30){$\mathbf j_3$}          \put(10,10){\line(1,1){10}}\put(10,30){\line(1,-1){10}}          \put(30,20){\line(1,1){10}}\put(30,20){\line(1,-1){10}}          \put(20,20){\line(1,0){10}}\put(22,25){$\mathbf j$}          \put(20,20){\circle*{3}}\put(30,20){\circle*{3}}\end{picture}\end{array}    &=& \sum_{\mathbf k}
 \left\{\begin{array}{ccc}
		{\mathbf  j_1}  & \mathbf j_2 & \mathbf k \\                      \mathbf j_3 & \mathbf j_4 &\mathbf  j                  \end{array}\right\}\begin{array}{c}\setlength{\unitlength}{1 pt}\begin{picture}(40,40)      \put( 0,0){$\mathbf j_1$}\put( 0,40){$\mathbf j_2$}      \put(35,0){$\mathbf j_4$}\put(35,40){$\mathbf j_3$}      \put(10,10){\line(1,1){10}}\put(10,40){\line(1,-1){10}}      \put(20,30){\line(1,1){10}}\put(20,20){\line(1,-1){10}}      \put(20,20){\line(0,1){10}}\put(22,22){$\mathbf k$}      \put(20,20){\circle*{3}}\put(20,30){\circle*{3}}\end{picture}\end{array}
 = \sum_{\mathbf k} \  c_{\mathbf k} \ i_{\mathbf k} \end{eqnarray}
On the other hand, in (\ref{4}) the simple intertwiner labelled by $j$ is in fact formed by a couple of intertwiners, one for the left and one for the right component of $SO(4)$, having the same spin. Namely 
\begin{eqnarray}
i_j &=& 
\begin{array}{c}\setlength{\unitlength}{1 pt}
\begin{picture}(50,40)          \put( 0,0){$j_1$}\put( 0,30){$j_2$}          \put(45,0){$j_4$}\put(45,30){$j_3$}          \put(10,10){\line(1,1){10}}\put(10,30){\line(1,-1){10}}          \put(30,20){\line(1,1){10}}\put(30,20){\line(1,-1){10}}          \put(20,20){\line(1,0){10}}\put(22,25){$j$}          \put(20,20){\circle*{3}}\put(30,20){\circle*{3}}\end{picture}\end{array}\begin{array}{c}\setlength{\unitlength}{1 pt}\begin{picture}(50,40)          \put( 0,0){$j_1$}\put( 0,30){$j_2$}          \put(45,0){$j_4$}\put(45,30){$j_3$}          \put(10,10){\line(1,1){10}}\put(10,30){\line(1,-1){10}}          \put(30,20){\line(1,1){10}}\put(30,20){\line(1,-1){10}}          \put(20,20){\line(1,0){10}}\put(22,25){$j$}          \put(20,20){\circle*{3}}\put(30,20){\circle*{3}}\end{picture}\end{array}    = 
\no &=& \sum_{k,l}
\left\{\begin{array}{ccc}                      j_1  & j_2 & k \\                      j_3 & j_4 & j                  \end{array}\right\}  \left\{\begin{array}{ccc}                      j_1  & j_2 & l \\                      j_3 & j_4 & j                  \end{array}\right\}\begin{array}{c}\setlength{\unitlength}{1 pt}\begin{picture}(40,40)      \put( 0,0){$j_1$}\put( 0,40){$j_2$}      \put(35,0){$j_4$}\put(35,40){$j_3$}      \put(10,10){\line(1,1){10}}\put(10,40){\line(1,-1){10}}      \put(20,30){\line(1,1){10}}\put(20,20){\line(1,-1){10}}      \put(20,20){\line(0,1){10}}\put(22,22){$k$}      \put(20,20){\circle*{3}}\put(20,30){\circle*{3}}\end{picture}\end{array}\begin{array}{c}\setlength{\unitlength}{1 pt}\begin{picture}(40,40)      \put( 0,0){$j_1$}\put( 0,40){$j_2$}      \put(35,0){$j_4$}\put(35,40){$j_3$}      \put(10,10){\line(1,1){10}}\put(10,40){\line(1,-1){10}}      \put(20,30){\line(1,1){10}}\put(20,20){\line(1,-1){10}}      \put(20,20){\line(0,1){10}}\put(22,22){$l$}      \put(20,20){\circle*{3}}\put(20,30){\circle*{3}}\end{picture}\end{array}
\label{eq:recTheorem}
\end{eqnarray}
where the first diagram represents the self-dual and the second diagram the anti--selfdual components of the representation.  Therefore in the 4d case a simple link in one pairing is 
equal to a sum including non simple links (when $k\ne l$) links in another pairing. That is 
\begin{equation}i_j = i_{(j,j)} = \sum_{k,l}\ c_{kl}\ i_{(k,l)}  \ne \sum_{k}\ c_{k}\ i_{k}
\label{eq:recTheorem2}
\end{equation}

The reason of the discrepancy between the linear structures in (\ref{eq:recTheorem}) and  (\ref{eq:recTheorem2}) is not entirely clear too us.  This discrepancy, on the other hand, does not affect the computations in the theory or the interpretation of the boundary states in the model we are considering.  The reason is that the only intertwiner appearing in the spinfoam sum is the Barrett-Crane intertwiner, which decomposes into simple virtual links for any pairing:
\begin{eqnarray}
i_{\rm BC}\  = \  \ Ê
\sum_{j}\ \  (2j+1)
\begin{array}{c}\setlength{\unitlength}{1 pt}
\begin{picture}(50,40)          \put(10,10){\line(1,1){10}}\put(10,30){\line(1,-1){10}}          \put(30,20){\line(1,1){10}}\put(30,20){\line(1,-1){10}}          \put(20,20){\line(1,0){10}}\put(22,25){$j$}          \put(20,20){\circle*{3}}\put(30,20){\circle*{3}}\end{picture}
\begin{picture}(50,40)          \put(10,10){\line(1,1){10}}\put(10,30){\line(1,-1){10}}          \put(30,20){\line(1,1){10}}\put(30,20){\line(1,-1){10}}          \put(20,20){\line(1,0){10}}\put(22,25){$j$}          \put(20,20){\circle*{3}}\put(30,20){\circle*{3}}\end{picture}\end{array}
=\ \ \ 
\sum_{k}\ \  (2k+1)
\begin{array}{c}\setlength{\unitlength}{1 pt}\begin{picture}(40,40)      \put(10,10){\line(1,1){10}}\put(10,40){\line(1,-1){10}}      \put(20,30){\line(1,1){10}}\put(20,20){\line(1,-1){10}}      \put(20,20){\line(0,1){10}}\put(22,22){$k$}      \put(20,20){\circle*{3}}\put(20,30){\circle*{3}}\end{picture}\begin{picture}(40,40)      \put(10,10){\line(1,1){10}}\put(10,40){\line(1,-1){10}}      \put(20,30){\line(1,1){10}}\put(20,20){\line(1,-1){10}}      \put(20,20){\line(0,1){10}}\put(22,22){$k$}      \put(20,20){\circle*{3}}\put(20,30){\circle*{3}}\end{picture}\end{array}
\end{eqnarray}
\noindent 
Therefore if we identify the intertwiner $i_{j_{tt'}}$ with the LQG intertwiner  $i_{\mathbf j_{tt'}}$, we obtain simply and consistently equation (\ref{pippo}). 

\section{Non-diagonal terms of the graviton operator}

The non--diagonal terms can be computed from 
\begin{equation}
E^{Ii}(n)E^{J}_i(n)|s\rangle = \hbar G\  |s_{nIJ}\rangle 
\end{equation}
where $s_{nIJ}$ is the spin network obtained by adding an infinitesimal link of color 1, joining  (``grasping") the links $I$ and $J$ infinitesimally close to the node $n$.   This is a spin network with the same graph and the same spins associated to the spins, but a different intertwiner at the node.  Assume for simplicity that the pairing chosen to label the intertwiner is such that the two grasped edges are paired together. Simple recouping algebra gives
\begin{equation}
E^{Ii}(n)E^{J}_i(n)|j\rangle 
= \hbar G\   \frac{{\rm Tet\!\left[\begin{array}{ccc}             j_1 & j_1 & j \\            j_2 & j_2 & 1    \end{array}\right]}}{\theta(j_1,j_2,j)}\ | j\rangle 
\end{equation}
where  and $j_1$and $j_2$ are the spins of the links $I$ and $J$. This gives 
(\cite{book}, pg 258)
\begin{equation}
E^{Ii}(n)E^{J}_i(n)|j\rangle 
= \hbar G\  \f12\left[j(j+1)-j_1(j_1+1)-j_2(j_2+1)\right] \ | j\rangle. 
\end{equation}

In the case in which the grasped links are not the paired ones, the action of the operator can simply be obtained by changing basis with the recouping theorem. This gives
\begin{equation}
E^{Ii}(n)E^{J}_i(n)(\vec{x})|j\rangle 
= \hbar G\ 
\sum_{k} c_j^k\ 
  |k\rangle 
\end{equation}
where
\begin{equation}
c_j^k= \hbar G\ 
\sum_{l}
  \left\{\begin{array}{ccc}                      j_1  & j_2 & l \\                      j_3 & j_4 & j                  \end{array}\right\}
  \frac{{\rm Tet\!\left[\begin{array}{ccc}             j_1 & j_1 & l \\            j_2 & j_2 & 1    \end{array}\right]}}{\theta(j_1,j_2,j)}\ 
  \left\{\begin{array}{ccc}                      j_1  & j_2 & k \\                      j_3 & j_4 & l                  \end{array}\right\}
\end{equation}
Here $j_1$ and $j_2$ are the spins of the two grasped links and 
$j_3$ and $j_4$ the other two. 

In conclusion
\begin{eqnarray}
\langle k | E^{Ii}(n)E^{J}_i(n) | j\rangle &=& 
 \hbar G\   j(j+1)\ \delta_{jk}
 \label{h1}
\end{eqnarray}
when $I=J$; here $j$ is the spin of the grasped link.
\begin{eqnarray}
\langle k|  E^{Ii}(n)E^{J}_i(n)| j\rangle &=& 
 \hbar G   \frac{{\rm Tet\!\left[\begin{array}{ccc}             j_1 & j_1 & j \\            j_2 & j_2 & 1    \end{array}\right]}}{\theta(j_1,j_2,j)} \delta_{jk}
 \label{h2}
\end{eqnarray}
for the grasping on paired links; here $j$ is the virtual link. And 
\begin{equation}
\langle k| E^{Ii}(n)E^{J}_i(n)| j\rangle =
 \hbar G\   \sum_{l}   
  \left\{\begin{array}{ccc}                      j_1  & j_2 & l \\                      j_3 & j_4 & j                  \end{array}\right\}
  \frac{{\rm Tet\!\left[\begin{array}{ccc}             j_1 & j_1 & l \\            j_2 & j_2 & 1    \end{array}\right]}}{\theta(j_1,j_2,j)}\ 
  \left\{\begin{array}{ccc}                      j_1  & j_2 & k \\                      j_3 & j_4 & l                  \end{array}\right\}
 \label{h3}
\end{equation}
for the grasping of unpaired links.  The action of $E^{ai}(n)E^{b}_i(n)$ is the immediate from
$E^{ai}(n) E^{b}_i(n) = n^a_I n^b_J E^{Ii}(n) E^{J}_i(n)$ where $n^a_I$ is the inverse matrix to $n_a^I$. 

The calculation of the non--diagonal terms of the propagator, involving the action of these operators, is in progress \cite{matrice}.

\end{document}